# Signal-to-noise and spatial resolution in in-line imaging.
# 2. Phase-contrast tomography

**Timur E. Gureyev[a*], David M. Paganin[b], Konstantin M. Pavlov[cbd], Anton Maksimenko[e] and Harry M. Quiney[a]**

[a]School of Physics, University of Melbourne, Parkville, Victoria, 3010, Australia
[b]School of Physics and Astronomy, Monash University, Clayton, Victoria, 3800, Australia
[c]School of Physical & Chemical Sciences, University of Canterbury, Christchurch, New Zealand
[d]School of Science and Technology, University of New England, Armidale, New South Wales, Australia
[e]Australian Synchrotron, Australian Nuclear Science and Technology Organisation, Sydney, New South Wales, Australia

Correspondence email: timur.gureyev@unimelb.edu.au

**Funding information** National Health and Medical Research Council (grant No. APP2011204).

**Abstract** In the first part of this paper, quantitative aspects of propagation-based phase-contrast imaging (PBI) were investigated using theoretical and numerical approaches, as well as experimental two-dimensional PBI images collected with plane monochromatic X-rays at a synchrotron beamline. In this second part, signal-to-noise ratio, spatial resolution and contrast are studied in connection with the radiation dose in three-dimensional PBI images of breast tissue samples obtained using propagation-based phase-contrast computed tomography (PB-CT) with energy-integrating and photon-counting detectors. The analysis is based on the theory of PBI and PB-CT using the homogeneous Transport of Intensity equation (Paganin's method). A biomedical image quality characteristic, suitable for quantitative assessment of X-ray images of biological samples, is introduced and applied. The key factors leading to high values of the biomedical imaging quality in PBI and to relatively low values of the same quality metric in CT imaging are identified and discussed in detail. This study is aimed primarily at developing tools for quantitative assessment and optimization of medical PB-CT imaging, initially at synchrotron facilities, with the prospect of subsequent transfer of the technology to medical clinics.

**Keywords: X-ray imaging, computed tomography, phase contrast, spatial resolution.**

#### 1. Introduction

In the first part of this paper (Gureyev et al., 2024), we presented theoretical results, numerical simulations and analysis of experimental data collected using propagation-based phase-contrast X-ray imaging (PBI) at a synchrotron source. Background information about PBI and its general advantages over conventional attenuation-based X-ray imaging were also discussed in the first part of this paper and therefore will not be repeated here. The theoretical approach used there, which is also used in the present second part of the paper, was largely based on the noise-resolution uncertainty (NRU) relationship (Gureyev et al., 2014, 2016, 2020; De Hoog et al., 2014). The NRU states that, for a broad class of linear imaging systems, the ratio of the squared signal-to-noise ratio (SNR) to the minimal spatially resolvable volume, $\Delta^n$ (where $\Delta$ is the spatial resolution and $n = 2$ for planar imaging systems, $n = 3$ in 3D imaging, etc.) depends only on the incident photon fluence. In particular, $\mathrm{SNR}^2 / \Delta^n$ remains unchanged after detector pixel binning, linear image filtering or denoising, etc. The invariance of this ratio under linear photon-number-conserving transformations of imaging systems is based on the fact that, at least for Poissonian photon-counting statistics, the ratio $\mathrm{SNR}^2 / \Delta^n$ corresponds to the mean number of detected photons that interacted with a minimal resolvable volume of the sample in the process of imaging. The latter quantity determines the amount of Shannon information that the system is capable of extracting about the imaged sample (Gureyev et al., 2016). As such, $\mathrm{SNR}^2 / \Delta^n$ is an essential intrinsic characteristic of the imaging system. It is, of course, trivial to decrease the ratio $\mathrm{SNR}^2 / \Delta^n$ by applying an SNR-reducing operation to an image, e.g. adding artificial noise to it. Note, however, that if an image processing operation utilises *a priori* information about the imaged sample, it can also potentially increase the ratio $\mathrm{SNR}^2 / \Delta^n$ without violating the NRU. In particular, this can take place in deep machine learning (artificial intelligence) based image denoising methods, such as those employing the highly successful UNet architecture (Ronneberger et al., 2015; Pakzad et al., 2025).

It has been argued (see e.g. Gureyev et al., 2017; Gureyev et al., 2024) that the key benefit of PBI is its ability to "beneficially violate" the NRU, i.e. increase $\mathrm{SNR}^2 / \Delta^n$ in comparison with an equivalent attenuation-based imaging setup (with zero propagation distance) without increasing the photon fluence or using *a priori* information. This can be achieved in PBI either by improving the spatial resolution without lowering the SNR (in the forward process of PBI image formation) or by increasing the SNR without spoiling the spatial resolution (in the combined process of forward PBI imaging and subsequent "phase retrieval" in accordance with Paganin's method (Paganin et al., 2002)). Since the ratio $\mathrm{SNR}^2 / \Delta^n$ can be increased this way in PBI, as demonstrated in multiple publications to date, see e.g. (Gureyev et al., 2014, 2024; Kitchen et al., 2017; Brombal et al., 2018), this seems to represent a true violation of the NRU. It may appear that additional information is created "out of nothing" in PBI, since the average number of photons per unit volume do not change in the process of forward free-space propagation of the beam. In fact, this phenomenon is rooted in the imaging physics, and more specifically, in the process of interaction of coherent radiation with an imaged sample. The NRU implicitly assumes that any information about the imaged sample can only be obtained by directly counting the photons that have been collected at the detector pixels. In that process, each photon carries no more than a single bit of information: essentially either the photon was transmitted through the sample and then detected (1), or it was absorbed or scattered away by the sample and hence not detected (0). However, in phase-contrast imaging, one can also detect and measure the phase shifts that the photons acquire upon coherent scattering by the atoms of the imaged sample. This information is made detectable in PBI by means of free-space propagation which allows the transmitted photons to interfere and thus reveal the phase information in the registered pattern of photon fluence. This detected signal is no longer binary, as in attenuation-based imaging, but can take any real value in the form of the measured phase shift. This means that the assumptions used in NRU no longer hold in phase-contrast imaging. The difference between the absorption and phase imaging here resembles the difference between the classical and quantum computing, with the former operating with binary bits and the latter utilizing qubits that can be in a weighted superposition of two states (Nielsen & Chuang, 2010).

An even more important factor leading to the "beneficial violation" of NRU and the "unreasonable" effectiveness of PBI (Gureyev et al., 2017) is the difference between the "coupling strengths" of the scattering channels corresponding to absorption and phase shifts in the interaction of incident photons with the imaged sample. In the case of imaging soft biological tissues with hard X-rays, the phase "coupling" is about three orders of magnitude stronger than the absorption "coupling" (Momose, 2020). Therefore, a phase-contrast method, like PBI, can overcome the limits set in NRU by utilising stronger scattering channels that are not taken into account in the theory underpinning NRU. The gain in SNR in PBI, in comparison with the equivalent absorption-based imaging at the same incident photon fluence and spatial resolution, was discussed in the first part of this paper and is further studied below in the case of three-dimensional (3D) imaging. It turns out that this gain is determined by the ratio of two dimensionless parameters, $\gamma / N_F$, where $\gamma$ is essentially the ratio of the strengths of the phase and absorption channels of scattering in the sample and $N_F$ is the "minimal Fresnel number" that is equal to $\Delta^2$ divided by the X-ray wavelength and by the sample-to-detector propagation distance. In biomedical PBI and propagation-based phase-contrast computed tomography (PB-CT), the ratio $\gamma / N_F$ can be of the order of 10 - 100. This effectively means that each detected photon in PB-CT can potentially carry hundreds of bits of information about the imaged sample, which appears to be an interesting and non-trivial phenomenon. It also leads to a potential decrease of the radiation dose of the order of $10^2$ - $10^5$ in PB-CT in comparison with conventional attenuation-based CT (Kitchen et al., 2017).

For the subject of the present paper it is important to consider another, this time "detrimental", violation of NRU which occurs in computed tomography (CT). It is well known that CT is a mathematically "ill posed" technique (Natterer, 2001). While "ill-posedness" of a system of equations can indicate non-existence or non-uniqueness of solutions, CT reconstruction actually does deliver a unique solution, i.e. a 3D reconstructed distribution, for any correctly sampled input projection data. However, CT is mathematically unstable, i.e. it amplifies the noise present in the input data. This can be understood by looking at the eigenvalues of the (forward) X-ray transform (consisting of X-ray projections collected at a set of view angles over 180 or 360 degrees), which decrease with the radial frequency of the corresponding eigenfunctions (Natterer, 2001). As a consequence, the X-ray transform progressively suppresses higher-order spatial frequencies in the input data. Accordingly, the eigenvalues of the inverse X-ray transform, i.e. the CT reconstruction operator, increase in magnitude with the radial frequency of its eigenfunctions. Therefore, CT reconstruction amplifies higher spatial frequencies more strongly than the lower ones. This can be also observed through the presence of the "ramp" filter in the continuous CT reconstruction (see the next Section for details). As a typical signal in CT decreases more rapidly with increasing spatial frequencies compared to a typical noise profile, this means that CT reconstruction usually amplifies the noise more than the signal, thus reducing the SNR. This leads to a detrimental violation of the NRU via the fact that $SNR^2$ in 3D CT is proportional to $\Delta^4$, rather than $\Delta^3$, as it should be in NRU (Howells et al., 2009; Nesterets & Gureyev, 2014). Nevertheless, when CT is combined with PBI, into the method known as PB-CT, it becomes significantly more efficient in terms of the increased $SNR^2 / \Delta^n$ ratio in the reconstructed data, in comparison with conventional attenuation-based CT. This was previously demonstrated both theoretically and experimentally (Nesterets & Gureyev, 2014; Kitchen et al., 2017; Brombal et al., 2020). Mathematically, the effectiveness of combining PBI with CT comes from the suppression of the noise-amplifying CT ramp filter by the noise-suppressing filter function of Paganin's method (Bronnikov, 1999; Paganin, 2002; Gureyev et al., 2006). As a result, $SNR^2$ in PB-CT becomes almost independent of the spatial resolution (Nesterets & Gureyev, 2014). This subject is also central to the present paper.

In order to make this second part of our paper largely self-contained, we have recalled the necessary basic facts about PB-CT in Section 2 below. In Section 3, we considered the question of quantitative assessment of performance of imaging systems in general and phase-contrast X-ray imaging of biological samples, in particular. Several of our previous studies, including the first part of this paper (Gureyev et al., 2024), utilized the "intrinsic imaging quality" characteristic, $Q_S$, the square of which is equal to the ratio of the quantity $SNR^2 / \Delta^n$ discussed above to the corresponding incident $n$-dimensional photon fluence, $I_{in}$, $Q_S^2 = SNR^2 / (I_{in} \Delta^n)$. The NRU is essentially equivalent to the statements that $Q_S$ cannot be larger than unity and that $Q_S$ is conserved in photon-number-conserving linear transformations of the imaging system. The intrinsic imaging quality characteristic corresponds to the amount of Shannon information that the imaging system is capable of extracting per single incident photon (Gureyev et al., 2016). It is typically measured in "flat fields", i.e. in images or parts of images obtained with no sample interaction with the beam, in which case $Q_S$ depends only on the properties of the imaging system alone. As in practice, especially in biomedical imaging, the image contrast generated by the sample and the radiation dose delivered to the sample in the process of imaging, are both critically important, it appears useful to consider a modified form of the intrinsic imaging characteristic that would explicitly include these two parameters into an image quality metric. We introduce such a metric, which we have tentatively named the "biomedical imaging quality characteristic", $Q_C$, in Section 4. In Section 5, we discuss the theoretical gain that PBI and PB-CT can deliver for $Q_S$ and $Q_C$, and then evaluate the gain in imaging quality of experimental images of breast tissue samples obtained using PB-CT in Section 6. This is followed by concluding remarks in Section 7.

A few technical derivations, the results of which are referenced in the paper, can be found in Appendices A-C. The main text of the paper also contains a relatively large number of mathematical expressions, which are essential to the considered problems and their solutions. For each mathematical expression or equation, we discuss both their physical meaning and their role in the physical picture and practice of PB-CT imaging in general and in biomedical imaging applications, in particular.

**2. Basic theory of propagation-based phase-contrast tomography**

Let $\exp(ikz)\Psi(\mathbf{r})$, $\Psi(\mathbf{r}) = I^{1/2}(\mathbf{r})\exp[i\varphi(\mathbf{r})]$, be the (slowly varying envelope of the) complex amplitude of a scalar monochromatic paraxial X-ray wave (beam) with wavenumber $k = 2\pi / \lambda$, where $\lambda$ is the wavelength, $\mathbf{r} = (\mathbf{r}_\perp, z)$ are Cartesian coordinates in three-dimensional (3D) space, $z$ is the beam propagation direction and $\mathbf{r}_\perp = (x, y)$ are the coordinates in transverse planes. The square modulus of the complex amplitude, $I(\mathbf{r})$, will be identified with the photon fluence and expressed in photon number per unit area. We assume that the X-ray wave near the vicinity of the $z$-origin of coordinates is a plane wave: $\Psi(\mathbf{r}_\perp, 0) = \Psi_{in} = I_{in}^{1/2}$. The process of transmission of this incident plane wave through a thin object with complex refractive index $n(\mathbf{r};\lambda) = 1 - \delta(\mathbf{r};\lambda) + i\beta(\mathbf{r};\lambda)$, located in the vicinity of the origin of coordinates immediately upstream of the "object plane", $z = 0$, can be described by the Beer-Lambert law

$$\Psi_0(\mathbf{r}_\perp, \theta) = \Psi_{in} \exp[(i/2)(\mathbf{P}n)(\mathbf{r}_\perp, \theta)], \tag{1a}$$

$$(\mathbf{P}f)(\mathbf{r}_\perp, \theta) = 2k \iint f(x', y, z')\delta_1\left(x - x'\sin\theta - z'\cos\theta\right)dx'dz', \tag{1b}$$

where $\Psi_0(\mathbf{r}_\perp, \theta)$ is the transmitted complex amplitude in the "object" plane, $z = 0$, $\delta_1(x)$ is the one-dimensional Dirac delta function and $\mathbf{P}$ is the X-ray transform (projection) operator (Natterer, 2001) which integrates its argument, $f$, along straight lines parallel to the beam direction. The primed Cartesian coordinates, $\mathbf{r}' = (x', y, z')$, in eq.(1b) are associated with the sample rotated by angle $\theta$ around the $y$ axis with respect to the fixed coordinates $\mathbf{r}$, which are associated with the fixed incident beam direction and the detector plane. Note that, starting from eqs.(1a)-(1b), we omit the dependence of any quantities on the wavelength for brevity. The propagation of the transmitted complex amplitude from the object plane $z = 0$ to the image (detector) plane $z = R$ is described by the Fresnel diffraction integral:

$$\Psi_R(\mathbf{r}_\perp, \theta) = (i\lambda R)^{-1} \iint \exp[ik \,|\, \mathbf{r}_\perp - \tilde{\mathbf{r}}_\perp |^2 / (2R)] \Psi_0(\tilde{\mathbf{r}}_\perp, \theta) d\tilde{\mathbf{r}}_\perp . \quad (2)$$

The assumption about the "thinness" of the object means that, for any rotational angle $\theta$, $\delta(\mathbf{r}') = \beta(\mathbf{r}') = 0$ outside the narrow slab $-L < z < 0$, where $0 < L << R$.

A complex amplitude is called monomorphous (or homogeneous) within some region of space (Paganin et al., 2002; Paganin et al., 2004), if the proportionality coefficient $\gamma$ between its phase and the logarithm of its modulus is independent of the position in that region. Monomorphous amplitudes arise e.g. in the object plane after transmission of an incident plane monochromatic X-ray wave through an object having the same chemical composition everywhere, possibly with a spatially-varying density. Indeed, for such objects, the ratio of the real decrement to the imaginary part of refractive index is the same at any point inside the sample: $\delta(\mathbf{r}) / \beta(\mathbf{r}) = \gamma = const$ (Paganin et al., 2004). Therefore, the phase, $\varphi_0(\mathbf{r}_\perp, \theta) = \varphi_{in} - (1/2)(\mathbf{P}\delta)(\mathbf{r}_\perp, \theta)$, and the square modulus, $I_0(\mathbf{r}_\perp, \theta) = I_{in} \exp[-(\mathbf{P}\beta)(\mathbf{r}_\perp, \theta)]$, of a wave transmitted through such an object satisfy a simple relationship:

$$\varphi_0(\mathbf{r}_\perp, \theta) - \varphi_{in} = (\gamma / 2) \ln[I_0(\mathbf{r}_\perp, \theta) / I_{in}] . \quad (3)$$

If the object-plane intensity also varies sufficiently slowly, such that $| a^2 \nabla_\perp^2 I_0(\mathbf{r}_\perp, \theta) | << I_0(\mathbf{r}_\perp, \theta)$, where $a \equiv \sqrt{\gamma R \lambda / (4\pi)}$ and $\nabla_\perp^2 = \partial_x^2 + \partial_y^2$ is the transverse 2D Laplacian, the square modulus of eq.(2) can be accurately approximated by the homogeneous Transport of Intensity equation (TIE-Hom) (Paganin et al., 2002):

$$I_R(\mathbf{r}_\perp, \theta) = (1 - a^2 \nabla_\perp^2) I_0(\mathbf{r}_\perp, \theta) = I_{in} (1 - a^2 \nabla_\perp^2) \exp[-(\mathbf{P}\beta)(\mathbf{r}_\perp, \theta)] . \quad (4)$$

Note that eq.(4) describes the propagation of the transmitted photon fluence from the object plane to the image plane without any explicit references to the phase distribution, i.e. without reference to the real part of the refractive index of the imaged object ("the sample"). Of course, in the case of non-trivial propagation distances $R$, this can only be possible due to the "homogeneity assumption", $\delta(\mathbf{r}) = \gamma\beta(\mathbf{r})$, about the sample, which allows for the phase information to be included in eq.(4) implicitly. Interestingly, eq.(4) has the structure of a finite-difference approximation to the diffusion equation, with $-a^2 / R$ playing the role of an $R$-dependent negative diffusion coefficient. Under this view, the use of the homogeneity assumption implies that it is now transverse intensity gradients (rather than phase gradients) that induce propagation-based transverse flow. The notion that intensity gradients drive flow is suggestive of diffusive energy transport, with the negative diffusion coefficient corresponding to sharpening rather than blur (cf. Gureyev et al., 2004).

It has been shown in (Thompson et al., 2019; Gureyev et al., 2022) that eq.(4) can be re-written as

$$C_R(\mathbf{r}_\perp,\theta) = [(1-a^2\nabla_\perp^2)\mathbf{P}\beta](\mathbf{r}_\perp,\theta) = [\mathbf{P}(1-a^2\nabla^2)\beta](\mathbf{r}_\perp,\theta), \qquad (5)$$

where $C_R(\mathbf{r}_\perp,\theta) \equiv 1 - I_R(\mathbf{r}_\perp,\theta)/I_{in} \cong -\ln[I_R(\mathbf{r}_\perp,\theta)/I_{in}]$ is the "PBI contrast function" and $\nabla^2 = \partial_x^2 + \partial_y^2 + \partial_z^2$ is the 3D Laplacian. In particular, eq.(5) effectively states that the TIE-Hom operator commutes with the projection operator. As a result, the "propagated" photon fluence at any rotational position of the sample can be obtained by applying the 3D TIE-Hom operator first, followed by the X-ray projection at the corresponding rotational position of the sample.

If a full set of propagated transmission images at different view angles in a 180 degree range, $\{I_R(\mathbf{r}_\perp,\theta): 0 \le \theta < \pi\}$, is available, eq.(5) can be inverted to reconstruct the 3D distribution of the refractive index $n(\mathbf{r}) = 1 + (i-\gamma)\beta(\mathbf{r})$ in a homogeneous sample (Gureyev et al., 2006; Thompson et al., 2019):

$$\beta(\mathbf{r}) = \mathbf{P}^{-1}(1-a^2\nabla_\perp^2)^{-1}C_R(\mathbf{r}_\perp,\theta) = (1-a^2\nabla^2)^{-1}\mathbf{P}^{-1}C_R(\mathbf{r}_\perp,\theta), \qquad (6)$$

where $\mathbf{P}^{-1}$ is the inverse X-ray transform (i.e. CT reconstruction) operator (Natterer, 2001), while $(1-a^2\nabla_\perp^2)^{-1}$ and $(1-a^2\nabla^2)^{-1}$ are the inverse TIE-Hom operators in 2D and 3D, respectively. Note that at $R=0$ we have $a=0$ and also $C_0(\mathbf{r}_\perp,\theta) \cong -\ln[I_0(\mathbf{r}_\perp,\theta)/I_{in}] = (\mathbf{P}\beta)(\mathbf{r}_\perp,\theta)$. Therefore, at $R=0$ eq.(6) transforms into the conventional CT reconstruction formula, $\beta(\mathbf{r}) = -\mathbf{P}^{-1}\ln[I_0(\mathbf{r}_\perp,\theta)/I_{in}]$ (Natterer, 2001). Of course, in practice, the infinite set of projections, $\{I_R(\mathbf{r}_\perp,\theta): 0 \le \theta < \pi\}$, is substituted by a properly sampled finite discrete set of measured fluence values, $\{I_R(\mathbf{r}_{\perp,lm},\theta_n): l = 1,2,...,L; m = 1,2,...,M; l = 1,2,...,N\}$, collected at detector pixel positions $(x_l, y_m)$ and view angles $\theta_n$ (Natterer, 2001). A properly sampled set of such projection values allows one to reconstruct a unique distribution of discretely sampled values of $\beta(\mathbf{r}_{lmn})$ at *LMN* voxels using a discretised version of eq.(6).

The inverse X-ray transform operator $\mathbf{P}^{-1}$ can be expressed either numerically, e.g. using a gridding-based implementation of the projection-slice theorem (Natterer, 2001; Gureyev et al., 2022), or analytically, e.g. with the help of the filtered back-projection (FBP) form of the inverse X-ray transform (Natterer, 2001):

$$\mathbf{P}^{-1}[g(\mathbf{r}_\perp,\theta)](\mathbf{r}) = (2k)^{-1}\int_0^{\pi}\int_{-\infty}^{\infty}\int_{-\infty}^{\infty}\exp\left\{i2\pi\left[\xi\left(x\sin\theta + z\cos\theta\right) + \eta y\right]\right\}|\xi|\hat{g}(\xi,\eta,\theta)d\xi d\eta d\theta, \quad (7)$$

where $\hat{f}(\mathbf{q}_\perp) = \iint \exp(-i2\pi\mathbf{q}_\perp \bullet \mathbf{r}_\perp)f(\mathbf{r}_\perp)d\mathbf{r}_\perp$ is the 2D Fourier transform and $\mathbf{q}_\perp = (\xi,\eta)$ is dual to $\mathbf{r}_\perp = (x, y)$.

The 2D and 3D TIE-Hom retrieval operators, $(1 - a^2\nabla_\perp^2)^{-1}$ and $(1 - a^2\nabla^2)^{-1}$, have simple analytical representations in Fourier space, where they can be expressed as multiplication by the functions $1/(1 + 4\pi^2 a^2 \rho_\perp^2)$ and $1/(1 + 4\pi^2 a^2 \rho^2)$, respectively (here and below we use the notation $\rho_\perp \equiv |\mathbf{k}_\perp|/(2\pi) = \sqrt{k_x^2 + k_y^2}/(2\pi)$ and $\rho \equiv |\mathbf{k}|/(2\pi) = \sqrt{k_x^2 + k_y^2 + k_z^2}/(2\pi)$). In real space, the 2D operator $(1 - a^2\nabla_\perp^2)^{-1}$ can be represented as a convolution with the function $T_{2,inv}(\mathbf{r}_\perp) \equiv K_0(r_\perp/a)/(2\pi a^2)$, where $K_0$ is the zero-order modified Bessel function of the second kind (Nesterets & Gureyev, 2014). The 3D operator $(1 - a^2\nabla^2)^{-1}$ can be expressed as a convolution with the Yukawa (screened Coulomb) potential $T_{3,inv}(\mathbf{r}) = (4\pi a^2 r)^{-1}\exp(-r/a)$ (Sakurai, 1967).

Equations (5) and (6) provide one-to-one forward and inverse mappings between (i) the 3D distributions of the imaginary part of refractive index in a homogeneous sample and (ii) the PBI contrast function collected in an imaging experiment with the sample scanned over 180 degrees of rotation around a fixed axis perpendicular to the X-ray beam direction and a detector located in the so-called "near-Fresnel" region, where $|a^2\nabla_\perp^2 I_0(\mathbf{r}_\perp,\theta)| << I_0(\mathbf{r}_\perp,\theta)$. The key question that we want to address in the present paper is that of the "gain" in the SNR-to-spatial-resolution ratio in the PB-CT reconstruction of $\beta(\mathbf{r})$ from the propagated projection images $\{I_R(\mathbf{r}_\perp,\theta): 0 \le \theta < \pi\}$, compared to the conventional CT reconstruction at $R = 0$, as a function of propagation distance $R$ and other relevant parameters.

#### 3. Quantitative performance measures of an imaging system

In the case of images collected with a position-sensitive detector, we shall distinguish between a stochastic registered distribution of the photon fluence, $I(\mathbf{r}_\perp)$, and its mean value, $\bar{I}(\mathbf{r}_\perp)$, which can be obtained by point-wise statistical averaging of an ensemble of equivalent images collected repeatedly under the same conditions. More generally, we consider random functions $I(\mathbf{r})$ in $n$-dimensional space, where, for example, $n = 2$ in the case of conventional planar images and $n = 3$ in the case of CT-reconstructed volumes.

To avoid extensive repetitions, we shall refer to some definitions included in the first part of the present paper (Gureyev et al., 2024). The SNR was defined in eq.(5) of (Gureyev et al., 2024) as the ratio of the mean of a random function $I$ at a point $\mathbf{r}$ to its standard deviation, $\text{SNR}[I](\mathbf{r}) = \bar{I}(\mathbf{r}) / \sigma_I(\mathbf{r})$. It was mentioned that, when the spatial ergodicity condition is satisfied, the signal and noise variance can be evaluated via spatial integrals over a flat vicinity of a given point.

The spatial resolution can be defined in terms of two different definitions of the width of the point-spread function (PSF) of the imaging system, $P(\mathbf{r})$:

(a) $\Delta[P]$, such that $\Delta^2[P] = (4\pi / n)\int |\mathbf{r}-\bar{\mathbf{r}}|^2\, P(\mathbf{r})d\mathbf{r} / \int P(\mathbf{r})d\mathbf{r}$, defined in eq.(6) of (Gureyev

et al., 2024), and

(b) $\tilde{\Delta}[P] = (\| P \|_1 / \| P \|_2)^{2/n}$, where $\| P \|_n \equiv (\int | P(\mathbf{r}) |^n \, d\mathbf{r})^{1/n}$, as defined in eq.(10) of (Gureyev et al., 2024). It was mentioned in (Gureyev et al., 2024) that $\Delta[P]$ and $\tilde{\Delta}[P]$ produce the same or very similar values for the width of many types of PSFs.

An important concept that was already discussed in (Gureyev et al., 2024) and will be used here, is that of the noise-resolution duality. Let $(I * O)(\mathbf{r})$ be a convolution of a random function $I(\mathbf{r})$ with a filter function $O(\mathbf{r})$. It was shown in (Gureyev *et al.*, 2016) that if $O(\mathbf{r})$ is much broader than $P(\mathbf{r})$, but varies much faster than $\bar{I}(\mathbf{r})$ and $\sigma_I^2(\mathbf{r})$, then the following ratio remains unchanged after the filtering:

$$\frac{SNR^2[O * I](\mathbf{r})}{\tilde{\Delta}^n[O * P]} = \frac{SNR^2[I](\mathbf{r})}{\tilde{\Delta}^n[P]} \, . \qquad (8)$$

In the case of 2D imaging with Poisson photon statistics, $\mathrm{SNR}^2$ is equal to the number of registered photons per detector pixel. Therefore, in this case, the “noise-resolution duality”, eq.(8), is just a restatement of a simple fact that larger “voxels” with the volume $\tilde{\Delta}^n[O * P]$, effectively created as a result of image filtering, contain more registered photons compared to the original “voxels” with the volume $\tilde{\Delta}^n[P]$, leading to the proportionally larger $\mathrm{SNR}^2$.

Now we can introduce the notion of (modified) intrinsic imaging quality characteristic, $\tilde{Q}_S$, of an imaging system, which is defined as

$$\tilde{Q}_S^2[I] \equiv \frac{SNR^2[I]}{\bar{I}_{in} \, \tilde{\Delta}^n[P]} , \qquad (9)$$

where $SNR^2[I]$ represents an average of $SNR^2[I](\mathbf{r})$ over a suitably representative flat area of an *n*-dimensional image, $\tilde{\Delta}[P]$ is the width of the PSF, as defined above, and $\bar{I}_{in}$ denotes the mean incident photon fluence, defined as the number of incident photons per *n*-dimensional volume. The subscript “S” in $\tilde{Q}_S$ serves as a reminder that this quantity is meant to be a characteristic of the imaging system, rather than a characteristic of an image. Note that $\tilde{Q}_S$ is a positive dimensionless quantity. Compared to the definition of intrinsic imaging quality characteristic, $Q_S$, used previously, e.g. in (Gureyev et *al*., 2014, 2016, 2024), eq.(9) utilises the spatial resolution $\tilde{\Delta}[P]$ instead of $\Delta[P]$. Although, as mentioned above, the two measures of the spatial resolution are identical or very similar for many popular functions *P*, $\tilde{Q}_S$ generally has more favourable formal mathematical properties compared to $Q_S$.

Note that $\bar{I}_{in} = \bar{n}_{vox} / \tilde{\Delta}^n_{vox}$, where $\bar{n}_{vox}$ is the mean number of photons incident on a single voxel during the imaging process and $\tilde{\Delta}^n_{vox}$ is the *n*-dimensional volume of the voxel. When the incident fluence is Poissonian, we have $\bar{n}_{vox} = SNR^2_{in}$, and hence

$$\tilde{Q}^2_S = p^n q, \tag{10}$$

where $p = \tilde{\Delta}_{vox} / \tilde{\Delta}[P]$ is the ratio of the linear size of the voxel to the spatial resolution, and $q = SNR^2 / SNR^2_{in} = DQE(0)$ is the detective quantum efficiency at zero frequency (Bezak et al., 2021). It has also been shown (Gureyev et *al*., 2016) that $\tilde{Q}_S$ corresponds to the Shannon information that the imaging system is able to extract from each incident photon. Therefore, imaging systems with higher $\tilde{Q}_S$ allow one to obtain better images that contain more information about the samples, compared to systems with lower $\tilde{Q}_S$.

It follows from eq.(8) that $\tilde{Q}_S$ remains unchanged in any post-detection linear filtering operation preserving the integral value of *I*, such as e.g. detector pixel binning, low-pass filtering or “phase retrieval” using the inverse TIE-Hom operators $(1-a^2\nabla^2_\perp)^{-1}$ and $(1-a^2\nabla^2)^{-1}$. In the first part of this paper, we discussed in detail that, although the forward TIE-Hom operator in eq.(4) is also linear and accurately describes the free-space propagation of the slowly varying mean fluence distribution, $\bar{I}(\mathbf{r}_\perp)$, from the object plane *z* = 0 to the detector plane *z* = *R*, it does not correctly describe the corresponding propagation of the noise term, $I(\mathbf{r}_\perp) - \bar{I}(\mathbf{r}_\perp)$ (we omit the projection angle $\theta$ here for brevity). According to the conservation of photon flux in free-space propagation, the noise variance in flat areas of images is the same at *z* = 0 and *z* = *R*, regardless of the propagation distance *R* (at least within the near-Fresnel region). On the other hand, the spatial resolution in TIE-Hom imaging improves after free-space propagation, so that $\tilde{\Delta}_R / \tilde{\Delta}_0 \cong \tilde{\Delta}_0 / (2\sqrt{\pi}\, a)$ in the 2D case (Gureyev et al., 2017). The increase of $\tilde{Q}_S[I] = SNR^2 / (\bar{I}_{in}\, \tilde{\Delta}^n)$ is then enabled by the improvement in the spatial resolution: $\tilde{Q}_S[I_R] \cong (\gamma / N_F)^{1/2}\, \tilde{Q}_S[I_0]$, where $N_F \equiv \tilde{\Delta}^2 / (\lambda R)$ is the "minimal" Fresnel number and $\gamma / N_F = 4\pi a^2 / \tilde{\Delta}_0^2$. As mentioned above, $\tilde{Q}_S$ remains unchanged after the TIE-Hom retrieval. Therefore, one obtains that $\tilde{Q}_S[I_{0,retr}] = (\gamma / N_F)^{1/2} \tilde{Q}_S[I_0]$, where $\tilde{Q}_S[I_{0,retr}]$ is the intrinsic imaging quality after the free-space propagation followed by the TIE-Hom retrieval (Gureyev et al., 2017). Since $\gamma / N_F$ can be much larger than one in hard X-ray imaging of biological samples, $\tilde{Q}_S[I_{0,retr}]$ can be much larger than $\tilde{Q}_S[I_0]$, and $\tilde{Q}_S[I_{0,retr}] = \tilde{Q}_S[I_R]$ can be significantly larger than one (Gureyev *et al.*, 2017; Gureyev *et al.*, 2024). Note that such large values of $\tilde{Q}_S$ only occur in imaging of "monomorphous" objects to which eq.(4) applies. We will investigate the corresponding behaviour of $\tilde{Q}_S$ in 3D, in the case of PB-CT, in Section 5 below.

#### 4. Characteristics of image quality dependent on the imaged sample

We have so far considered different performance metrics of imaging systems (SNR, $\tilde{\Delta}$, $\tilde{Q}_S$) without making reference to imaged samples. In practice, it is often required to analyse the image quality and optimise the performance of an imaging system for a particular class of samples. In this respect, one such metric that has not been considered above, is the image contrast. Indeed, useful image contrast can typically appear only in the presence of an imaged sample, and the strength of contrast inevitably depends on the sample, as well as on the imaging system. The contrast is often defined as $C = |\bar{I}_1 - \bar{I}_2| / (\bar{I}_1 + \bar{I}_2)$, where $\bar{I}_1$ and $\bar{I}_2$ are the mean pixel values in two adjacent areas of the image. We will use a slightly modified version of this definition, $C_m = |\bar{I}_1 - \bar{I}_2| / \max\{\bar{I}_1, \bar{I}_2\}$, which has similar properties to *C*. Indeed, both *C* and $C_m$ can take values between 0 and 1, both reach the maximum value of 1, when one of the two intensities is equal to zero, and the minimum value of 0, when the two intensities are equal. Moreover, in the case where $\bar{I}_2 = \bar{I}_1 / 2$, we get $C = 1/3$ and $C_m = 1/2$, with the latter result possibly being more natural on an intuitive level. We can then define the contrast to noise ratio as

$$CNR[I_1, I_2] = C_m \times SNR = \frac{|\bar{I}_1 - \bar{I}_2|}{\sigma_{I_m}}, \tag{11}$$

where $SNR = \bar{I}_m / \sigma_{I_m}$ is measured in the ("background") area with higher pixel values, $\bar{I}_m = \max\{\bar{I}_1, \bar{I}_2\}$. The CNR is an important characteristic of an image, and the goal is typically to maximize it. Alternatively, when the CNR is close to zero or is below the noise level, the image is likely to be considered useless.

One type of contrast which is often used in X-ray imaging is that between an absorbing sample and the flat-field background corresponding to the incident illumination. In this case we have $I_2 = I_{in}$ and $I_1 = I_{in}\exp[-\int \mu(\mathbf{r}_\perp, z)dz]$, where $\mu = 4\pi\beta / \lambda$ is the linear attenuation coefficient of the sample, the integral is along the rays passing through the sample, and the relevant mean values can be calculated over the transverse coordinates in the region of interest (ROI) in the image. When the absorption is weak, such that $\int \mu(\mathbf{r}_\perp, z)dz << 1$ and hence $\exp[-\int \mu(\mathbf{r}_\perp, z)dz \cong 1 - \int \mu(\mathbf{r}_\perp, z)dz$, the contrast is approximately equal to $C_m \cong \int \mu(\mathbf{r}_\perp, z)dz$.

In the case of PBI and PB-CT, an important model case of image contrast is represented by an edge of a slab of weakly absorbing ("transparent") material with the real decrement of refractive index, $\delta$. In the case of a parallel monochromatic X-ray beam, the corresponding phase shift is $\Delta\varphi = -(2\pi / \lambda)\int \delta(\mathbf{r}_\perp, z)dz$. The "edge contrast", representing a simple case of propagation-based phase contrast, is defined as the difference between the maximum and the minimum of pixel values in the first Fresnel fringe generated by the edge at the sample-to-detector distance *R*, divided by the sum of the same two extreme intensities. It has been shown (Gureyev et *al*., 2008) that such phase edge contrast can be expressed as $C_{m,PBI} \cong |\Delta\varphi| \lambda R / \tilde{\Delta}^2 = 2\pi R \tilde{\Delta}^{-2} \int \delta(\mathbf{r}_\perp, z)dz$. The latter expression is valid under the condition $N_F = \tilde{\Delta}^2 / (\lambda R) >> \max\{1, |\Delta\varphi|\}$, in the so-called near-Fresnel region.

Another important characteristic of an image is the radiation dose that has been delivered to the sample in the process of acquiring the image. The absorbed dose, $D_{ab}$, is the energy per unit mass absorbed by the sample, which is expressed in Gy = J / kg. The importance of the radiation dose can be due to the carcinogenic effect of ionising radiation in medical imaging, or due to the damage by electrons used for imaging in high-resolution electron microscopy, or for other reasons relevant to a particular imaging context. The absorbed dose is closely related to the kerma, $K$, which is the energy transferred from photons to kinetic energy of electrons in the unit mass of matter in which the kerma is measured:

$$D_{ab} = K(1-g), \tag{12}$$

where the coefficient $g$ is the fraction of energy lost to Bremsstrahlung and other radiative processes (Bezak et al., 2021). Both the absorbed dose and the kerma can be factorized into the product of the incident photon fluence, the mass absorption coefficient, $\mu / \rho$, where $\rho$ is the sample density, and the average energy absorbed, $\bar{E}_{ab}$, or transferred, $\bar{E}_{tr}$, respectively, at each interaction: $D_{ab} = \bar{I}_{in}\bar{R}_{ab}$, $\bar{R}_{ab} = (\mu / \rho)\bar{E}_{ab}$ and $K = \bar{I}_{in}\bar{R}_{tr}$, $\bar{R}_{tr} = (\mu / \rho)\bar{E}_{tr}$.

In the context of X-ray imaging of the breast, which is the most relevant to the experimental results analysed below, the standard measure of the dose is the mean glandular dose (MGD), $\bar{D}_g$, which is calculated from the measurements of the entrance air kerma, $K_{air} = \bar{I}_{in}R_{tr,air}$, according to the formula

$$\bar{D}_g = D_g^N K_{air}, \tag{13}$$

where $D_g^N$ is a dimensionless conversion factor which depends on the X-ray energy, the breast thickness and composition (such as glandularity) (Bezak et al., 2021). This conversion factor is usually obtained by Monte-Carlo simulations with breast-equivalent phantoms (Johns & Yaffe, 1985; Nesterets et *al.*, 2015). The maximum permissible MGD for a mammographic image is currently around 1 mGy, with some variation in the standards between different countries (Liu et al., 2022).

In order to include the effect of the sample on imaging quality, we first multiply the intrinsic imaging quality characteristic $\tilde{Q}_S$ by the contrast, $C_m$, which can be calculated for a feature of interest in the image. Then, in order to take into account the detrimental effect of the radiation dose, in accordance with the "linear no-threshold model" (Bezak et al., 2021), we also divide the new characteristic by the absorbed dose or by the MGD, depending on the context. However, the resultant quantity would depend on the measurement units, such as kg / J or g / calorie. In order to avoid this cumbersome dimensionality and make the new imaging quality metric dimensionless, we can follow the general idea behind the Hounsfield units and normalise our metric accordingly. It appears that, in the case of biomedical X-ray imaging, a natural normalization can be achieved with respect to X-ray absorption in air:

$$\tilde{Q}_C^2 = \tilde{Q}_S^2 C_m^2 (\bar{R}_{ab,air} / \bar{R}_{ab,tissue}) , \tag{14}$$

or, in the special case of breast imaging,

$$\tilde{Q}_C^2 = \tilde{Q}_S^2 C_m^2 (\bar{R}_{ab,air} / D_g^N \bar{R}_{tr,air}) \cong \tilde{Q}_S^2 C_m^2 / D_g^N . \tag{14a}$$

Here we used the fact that at X-ray energies typical for breast imaging (~20-40 keV), the coefficient $g_{air}$ is much smaller than unity (see e.g. Kato, 2014). We propose to call the metric $\tilde{Q}_C$ "biomedical imaging quality". Note that the word "intrinsic" is no longer relevant here, since, unlike $\tilde{Q}_S$, the new metric $\tilde{Q}_C$ substantially depends both on the imaging system and on the imaged sample. We used the normalization coefficient $\bar{R}_{ab,air}$ related to air in eqs.(14)-(14a), and not e.g. to water as in the case of Hounsfield units, because the sample is assumed to be imaged in air. In the case of breast imaging, the conversion factor $D_g^N$ for MGD is also defined with respect to the air kerma. If the sample is absent and there is only air, then eq.(14) and eq.(14a) both convert simply to $\tilde{Q}_C = \tilde{Q}_S C_m$. The latter quantity is likely to be equal to zero, since in a well-designed imaging system, the contrast should be close to zero in the absence of a sample.

Let us consider first the case of 2D imaging:

$$\tilde{Q}_{C,2D}^2 = \frac{C_m^2 \, SNR^2 \, \bar{R}_{ab,air}}{\bar{R}_{ab,tissue} \bar{I}_{in} \, \tilde{\Delta}^2} = \frac{CNR^2 \, \bar{R}_{ab,air}}{\bar{D}_{ab} \, \tilde{\Delta}^2} , \tag{15}$$

or in the case of breast imaging:

$$\tilde{Q}_{C,2D}^2 = \frac{C_m^2\, SNR^2\, \bar{R}_{ab,air}}{D_g^N\, \bar{R}_{tr,air} \bar{I}_{in}\, \tilde{\Delta}^2} = \frac{CNR^2\, \bar{R}_{ab,air}}{\bar{D}_g\, \tilde{\Delta}^2}\,. \tag{15a}$$

The mean absorbed dose (MAD) or MGD are typically measured with the help of ionisation chambers or dosimeters during an experiment, while CNR and $\tilde{\Delta}$ can be measured later in the collected images. In order to calculate the imaging quality metric $\tilde{Q}_{C,2D}$, one also needs to know the normalization constant $\bar{R}_{ab,air}$. In the case of monochromatic X-rays, we have $\bar{R}_{ab} = (\mu_{en} / \rho) E_{ph}$, where $\mu_{en} / \rho$ is the mass energy-absorption coefficient and $E_{ph} = hc / \lambda$ is the energy of a photon (Hubbell & Seltzer, 1996). Calculating the photon energy is trivial and the mass energy-absorption coefficient of air at a given X-ray energy can be found e.g. in NIST databases (Higgins et *al.*, 1992; Hubbell & Seltzer, 2004) and elsewhere. For example, at 32 keV,
$hc / \lambda \cong 6.626 \times 10^{-34} (\mathrm{kg\, m^2/s}) \times 2.998 \times 10^{8} (\mathrm{m/s}) / (3.875 \times 10^{-11} \mathrm{m}) \cong 5.127 \times 10^{-15}\, \mathrm{J}$,
$\mu_{en,air} / \rho_{air} \cong 0.01366\, \mathrm{m^2/kg}$ for dry air at sea level (Kato, 2014), and hence
$\bar{R}_{ab,air} \cong 7.003 \times 10^{-17}\, \mathrm{m^4 / s^2}$. In a typical medical X-ray imaging setup, $\bar{D}_{ab}$ and $\bar{D}_g$ can be of the order of $10^{-3}$ Gy and $\tilde{\Delta}$ can be of the order of 100 μm. In such cases, the dimensionless ratios $\bar{R}_{ab,air} / (\bar{D}_{ab} \tilde{\Delta}^2)$ and $\bar{R}_{ab,air} / (\bar{D}_g \tilde{\Delta}^2)$ will be of the order of $10^{-6}$. If the contrast is not too small, one will need to have $SNR^2$ of the order of $10^6$, or about 1000 detected photons per pixel, for $\tilde{Q}_{C,2D}$ to approach unity. Of course, any increase in SNR via an increased incident fluence will be associated with a simultaneous increase in the dose, so improving the biomedical imaging quality characteristic of an imaging setup is not a trivial task.

While eq.(9) for $\tilde{Q}_S$ has the same mathematical form for any *n* in *n*-dimensional imaging, eqs.(15) and (15a) do not retain the same form for $n \neq 2$, due to the intrinsically 2D nature of the definition of the dose. Let us see how eq.(15) changes in a 3D-imaging case, such as CT imaging. Using eq.(14), we obtain:

$$\tilde{Q}_{C,3D}^2 = \frac{C_m^2 \bar{R}_{ab,air}}{\bar{R}_{ab,tissue}} \frac{SNR_{3D}^2}{\bar{I}_{in,3D}\, \tilde{\Delta}^3}\,. \tag{16}$$

In order to correctly re-write the denominator of the right-hand side of eq.(16) in terms of the 3D dose, let us first express the sample volume as $V = \Omega L$, where $\Omega$ is the illuminated "front" surface area and $L \equiv V / \Omega$ is the effective depth of the sample. In the situation most frequently encountered in CT, the reconstructed sample volume is cylindrical: $V = \pi R_C^2 H$, where $R_C$ is the radius of the cylinder and $H$ is its height. The air kerma $K = \bar{I}_{in}\bar{R}_{tr}$ is typically measured with respect to the fluence $\bar{I}_{in}$ corresponding to a flat entrance surface with the area $\Omega = 2R_C H$ and hence $L = V / \Omega = (\pi / 2)R_C$ in this case. Let $N_{3D} = N_{2D}M_a$ be the total number of photons, or more precisely, the number of noise equivalent quanta, NEQ (Bezak et al., 2021), collected in the detector area corresponding to $\Omega$ during a CT scan. Here $N_{2D}$ is the average total number of NEQ*s* detected in each 2D projection and $M_a$ is the number of angles at which the CT projections have been acquired. Then, assuming that the measurements are performed in a uniform area of the images, $\bar{I}_{in,3D} = N_{3D} / V = N_{2D}M_a / (\Omega L) = \bar{I}_{in,2D}M_a / L$ and hence the denominator of eq.(16) is equal to $\bar{R}_{ab,tissue}\bar{I}_{in,3D}\tilde{\Delta}^3 = \bar{R}_{ab,tissue}\bar{I}_{in,2D}M_a\tilde{\Delta}^3 / L = \bar{D}_{ab,3D}\tilde{\Delta}^3 / L$, where $\bar{D}_{ab,3D} = \bar{D}_{ab,2D}M_a = \bar{R}_{ab,tissue}\bar{I}_{in,2D}M_a$ is the total MAD accumulated during the scan. Equation (16) can now be written as

$$\tilde{Q}_{C,3D}^2 = \frac{CNR_{3D}^2\bar{R}_{ab,air}L}{\bar{D}_{ab,3D}\tilde{\Delta}^3}. \qquad (17)$$

Similarly, in the case of breast imaging, we obtain:

$$\tilde{Q}_{C,3D}^2 = \frac{CNR_{3D}^2\bar{R}_{ab,air}L}{\bar{D}_{g,3D}\tilde{\Delta}^3}. \qquad (17a)$$

Note that the appearance of the factor $L$ in eqs. (17) and (17a) is the consequence of the fact that MAD and MGD are defined with respect to the 2D photon fluence, rather than the 3D photon fluence which appears in the 3D version of $\tilde{Q}_S^2$. In a limit case, when the number of projections is equal to one, one has $D_{ab,3D} = D_{ab,2D}$ and $D_{g,3D} = D_{g,2D}$, the voxel volume is $\tilde{\Delta}^3 = \tilde{\Delta}^2 L$, $SNR_{3D}^2 = SNR_{2D}^2$, and hence eq.(17) and eq.(17a) morph into eq. (15) and eq.(15a), respectively. Note also that, as can be seen in the latter example, the resolution volume $\tilde{\Delta}^3$ in eq.(17) and eq.(17a) may not be equal, in general, to $(\tilde{\Delta}^2)^{3/2}$, where $\tilde{\Delta}^2$ is the resolution area used in eq.(15) and eq.(15a). Therefore, $\tilde{\Delta}^2$ and $\tilde{\Delta}^3$ are just shorthand notations for $(\tilde{\Delta}_{2D})^2$ and $(\tilde{\Delta}_{3D})^3$, where $\tilde{\Delta}_{2D}$ and $\tilde{\Delta}_{3D}$ are the spatial resolutions in the 2D and 3D imaging, respectively.

It is important to remember that both $\tilde{Q}_S$ and $\tilde{Q}_C$ have been defined as imaging quality measures per single incident photon. In some cases, it can be useful to evaluate the quality of a particular 2D or 3D image, obtained with a given incident fluence or a given dose. Natural metrics for this purpose are represented by the product of the above metrics, defined per single photon, and the total number of the incident photons, $N_{tot}$, that was used to create the image, i.e. $\tilde{Q}_S N_{tot}$ and $\tilde{Q}_C N_{tot}$. Note that $\bar{I}_{in} = N_{tot} / V$, where $V$ is the $n$-dimensional volume of the imaged sample. Therefore, $N_{tot} / (\bar{I}_{in} \tilde{\Delta}^n) = V / \tilde{\Delta}^n = M_{vox}$, where $M_{vox}$ is the total number of $n$-dimensional voxels, and hence

$$N_{tot} \tilde{Q}_S^2 = M_{vox} SNR^2 = SNR_{tot}^2, \tag{18}$$

where $SNR_{tot}^2$ is the squared "total" SNR of the image, obtained by summing all the squared SNRs measured in individual voxels. The latter quantity is equal to the total number of noise-equivalent quanta, $NEQ_{tot}$, in the image (Bezak et al., 2021). Therefore, the intrinsic image quality, $N_{tot} \tilde{Q}_S$, provides a direct quantitative measure of the (Shannon) information content of the image. For comparison, recall, that $\tilde{Q}_S$ is closely related to Shannon's channel information capacity of the imaging system (Gureyev et al., 2016), which is the amount of information per single photon.

Similarly,

$$N_{tot} \tilde{Q}_C^2 = (C_m^2 \bar{R}_{ab,air} / \bar{R}_{ab,tissue}) N_{tot} \tilde{Q}_S^2 = C_m^2 SNR_{tot}^2 (\bar{R}_{ab,air} / \bar{R}_{ab,tissue}), \tag{19}$$

i.e. the square of the intrinsic quality of an image of a sample is equal to the total number of *NEQs* in the image, multiplied by the square of the contrast and by the ratio of the mass energy-absorption coefficients of air and the tissue. Note that, when using eq.(19), the contrast should be approximately uniform in the whole image. Alternatively, an appropriate kind of average contrast across the image can be used, or the metric from eq.(19) could be used for a particular part of the image, etc.

#### 5. Noise and spatial resolution in PB-CT

We proceed with the evaluation of the ratio $SNR^2 / \tilde{\Delta}^3$ in the distribution of $\beta(\mathbf{r})$ reconstructed from PB-CT scans collected in an image plane $z = R$, in comparison to the same ratio obtained in conventional CT scans collected in the object plane $z = 0$. The (average) "gain factor" in the intrinsic imaging quality of the PB-CT imaging, compared to the conventional CT, can be defined as follows (Nesterets & Gureyev 2014; Gureyev et al., 2017; Gureyev et al., 2024):

$$\tilde{G}_3 \equiv \frac{\tilde{Q}_{S,retr}}{\tilde{Q}_{S,0}} = \left( \frac{SNR[\beta_{retr}]}{\bar{I}_{R,in}^{1/2}\, \tilde{\Delta}^{3/2}[P_{retr}]} \right) \Big/ \left( \frac{SNR[\beta_0]}{\bar{I}_{0,in}^{1/2}\, \tilde{\Delta}^{3/2}[P_0]} \right), \qquad (20)$$

where the subindexes "0" and "*retr*" correspond to quantities obtained from the data collected by conventional CT at $z = 0$ and by PB-CT with TIE-Hom retrieval from the data collected at $z = R$, respectively, with $\bar{I}_{0,in}$ and $\bar{I}_{R,in}$ denoting the mean incident fluence in the object and the image planes.

If we consider the gain factor at a certain fixed level of the incident X-ray fluence, then, as discussed above, the mean fluences at $z = 0$ and $z = R$ are going to be equal, due to the conservation of photon flux in free-space propagation (ignoring the absorption in air, for simplicity). Moreover, if the TIE-Hom retrieval is performed with $a = \sqrt{\delta R \lambda / (4\pi\beta)}$, then, according to eq.(6), the same distribution of $\bar{\beta}$ is obtained after the reconstruction, regardless of the distance $R$. Indeed, the mean values of the input to eq.(6), $\bar{C}_R(\mathbf{r}_\perp, \theta)$, are the same in flat areas of images at different $R$ within the near-Fresnel region, due to the conservation of photon flux in free-space propagation. Furthermore, the inverse X-ray transform, $\mathbf{P}^{-1}$, does not depend on $R$, while the inverse TIE-Hom operators, $(1 - a^2\nabla_\perp^2)^{-1}$ and $(1 - a^2\nabla^2)^{-1}$, both preserve the mean value of their input for any $R$, since they can be represented as convolutions with filter functions $T_{2,inv}(\mathbf{r}_\perp)$ and $T_{3,inv}(\mathbf{r})$ which both have unit integrals. Therefore, both the signals and the spatial resolutions are the same in the case of conventional and PB-CT reconstructions, regardless of the value of *R*, i.e. $\bar{\beta}_{R,retr}(\mathbf{r}) = \bar{\beta}_0(\mathbf{r})$ and $\tilde{\Delta}[P_{retr}] = \tilde{\Delta}[P_0]$. In this case, eq.(20) reduces to a simpler expression, $\tilde{G}_3 = SNR[\beta_{retr}] / SNR[\beta_0] = \sigma_{\beta_0} / \sigma_{\beta_{retr}}$ (Nesterets & Gureyev, 2014).

Note that although eq.(20) is expressed in terms of the intrinsic imaging quality characteristic, the gain factor $\tilde{G}_3$ still depends on the type of imaged samples. Indeed, the ratio $\delta / \beta$ used in the TIE-Hom retrieval step is specific to a particular class of sample, and only for such samples can the gain in SNR can be achieved without a loss of spatial resolution after the TIE-Hom retrieval of the images collected at $z = R$ using the specified parameter $a = \sqrt{\delta R \lambda / (4\pi\beta)}$. Therefore, it is logical to take other relevant sample-specific factors, such as contrast and dose, into account and consider the gain coefficient for the biomedical imaging quality, $\tilde{Q}_C$. Since the distribution of the mean values $\bar{\beta}(\mathbf{r})$, obtained as a result of a PB-CT reconstruction, is invariant with respect to the propagation distance *R*, the contrast $C_m = |\bar{I}_1 - \bar{I}_2| / \max\{\bar{I}_1, \bar{I}_2\}$ in the reconstructed images, e.g. between the mean values of $\beta$ in adjacent areas of adipose and glandular breast tissues, is independent of *R*. Therefore, the improvement in the biomedical imaging quality characteristic $\tilde{Q}_C$ upon free-space propagation followed by the TIE-Hom retrieval, compared to the conventional CT at $R = 0$, is the same as in eq.(20):

$$\tilde{G}_3 = \frac{\tilde{Q}_{C,retr}}{\tilde{Q}_{C,0}} = \left( \frac{CNR[\beta_{retr}]}{\bar{D}_{ab}^{1/2} \, \tilde{\Delta}^{3/2}[P_{retr}]} \right) \Big/ \left( \frac{CNR[\beta_0]}{\bar{D}_{ab}^{1/2} \, \tilde{\Delta}^{3/2}[P_0]} \right). \qquad (21)$$

In the case of breast imaging, the MADs in eq.(21) can be replaced by the corresponding MGDs.

Furthermore, as mentioned in the first part of this paper (Gureyev et al., 2024), when imaging a class of samples with the same $\delta / \beta$ ratio, the gain factor is independent of the parameter $a = \sqrt{\delta R \lambda / (4\pi\beta)}$ used in the TIE-Hom retrieval, as long as this procedure satisfies the conditions required for validity of eq.(8). With smaller values of parameter *a*, the reconstructed images will have lower SNR, but higher spatial resolution, and the opposite will be true for larger values of $a$. The invariance of the gain factor with respect to the choice of parameter $a$ is important, because TIE-Hom retrieval in PB-CT imaging is often deliberately performed with smaller values of $a$ in order to retain some image-sharpening effects of coherent free-space propagation (Tavakoli Taba et al., 2019). Another reason for using a smaller value of $a$ in TIE-Hom retrieval is that the penumbral effect of the source contributes to the spatial resolution and already does some "phase-retrieval blur" (Beltran et al., 2018).

While both the signal and the spatial resolutions are the same in the case of conventional and PB-CT reconstructions, regardless of the value of $R$, the noise variance in the reconstructed distribution may vary dramatically as a function of $R$ in PB-CT at a fixed level of the incident fluence (Nesterets & Gureyev, 2014; Kitchen et al., 2017; Brombal et al., 2021). Let us consider the latter effect quantitatively, on the basis of the noise-resolution duality, eq.(8). According to the last expression in eq.(6), conventional CT and PB-CT differ only by the free-space propagation and the application of the 3D TIE-Hom retrieval operator, $(1-a^2\nabla^2)^{-1}$. As mentioned above, the action of the operator $(1-a^2\nabla^2)^{-1}$ can be expressed as a convolution with the function $T_{3,inv}(\mathbf{r})$. This fact is used in Appendix A to derive the following approximation for the 3D gain factor:

$$G_3 \cong \left( \frac{8\pi a^3 f(A)}{\tilde{\Delta}_{CT}^3} \right)^{1/2} \cong \frac{(\gamma R\lambda)^{3/4} f^{1/2}(A)}{\pi^{1/4} \tilde{\Delta}_{CT}^{3/2}} = \frac{f^{1/2}\left(\sqrt{\pi\gamma / (4N_{F,CT})}\right)}{\pi^{1/4}} \left( \frac{\gamma}{N_{F,CT}} \right)^{3/4}, \qquad (22)$$

where $f(A) = (\pi/2)[\arctan(A) - A/(1+A^2)]^{-1}$, $A = \pi a / \tilde{\Delta}_{CT}$ and $N_{F,CT} \equiv (\tilde{\Delta}_{CT}^3)^{2/3} / (R\lambda)$ is the Fresnel number corresponding to the spatial resolution after the CT reconstruction. Equation (22) was obtained under the condition that the width of $T_{3,inv}$ is much larger than the detector resolution, hence $8\pi a^3 >> \tilde{\Delta}_{CT}^3$, $A >> 1$ and $f(A) \sim 1$, implying that $G_3 > 1$. For completeness, we note that $G_1 = (4/\pi)^{1/4}(\gamma / N_F)^{1/4}$ and $G_2 = (\gamma / N_F)^{1/2}$ (Nesterets & Gureyev, 2014; Gureyev et *al*., 2017, 2024), where $N_F \equiv \tilde{\Delta}^2 / (R\lambda)$ and $\tilde{\Delta} \equiv \tilde{\Delta}[P]$ is the spatial resolution in the projection images.

While eq.(22) was obtained using the last equality in eq.(6), a somewhat different approximation for the 3D gain factor was obtained earlier on the basis the first part of eq.(6) (with 2D TIE-Hom retrieval) (Nesterets & Gureyev, 2014):

$$G_3 \cong \left( \frac{\pi / 6}{\ln(\gamma / N_F) - 1} \right)^{1/2} \frac{\gamma}{N_F} \quad . \tag{23}$$

A brief derivation of eq.(23) is given in Appendix B for completeness. This form of the gain factor was previously investigated and discussed (Kitchen et al., 2017; Brombal et al., 2018, Brombal, 2020). We shall see in the next section how well these theoretical estimates of the gain factor agree with our experimental results.

It follows from the analytical expression for the biomedical imaging quality of CT derived in Appendix C that

$$\tilde{Q}^2_{C,3D}[\beta_{retr}] \cong G_3^2 \frac{12 C_m^2 (\bar{\mu} L)^2 \exp(-\bar{\mu} L)}{\pi^2 D_g^N M_R} , \tag{24}$$

where $\bar{\mu}$ is the average linear attenuation coefficient and $M_R = L / \tilde{\Delta} = \pi R_C / (2\tilde{\Delta})$ is $\pi / 2$ times the number of spatial resolution units in the radius of the CT reconstruction cylinder. An analytical expression for conventional CT, i.e. $\tilde{Q}^2_{C,3D}[\beta_0]$, is obtained from eq.(24) when $G_3 = 1$. The constant factor $(12 / \pi^2)$ in eq.(24) corresponds to a particular implementation of the CT reconstruction, namely the filtered back-projection with a nearest neighbour interpolation; it can be slightly different for other implementations of the CT reconstruction (Nesterets & Gureyev, 2014). As could be expected, the biomedical imaging quality in eq.(24) is directly proportional to the image contrast, $C_m$, while being inversely proportional to the square root of the ratio of the MGD to the air kerma, $D_g^N = \bar{D}_g / K_{air}$. The term $(\bar{\mu} L)^2 \exp(-\bar{\mu} L)$ in the numerator of eq.(24) achieves the maximum value of approximately 0.54 at $\bar{\mu} L = 2$. The factor $M_R$ in the denominator of eq.(24) is a “proxy” for the highest radial frequency present in the reconstructed distribution of $\beta$, which determines the degree of ill-posedness of the CT reconstruction, as mentioned earlier (Natterer, 2001). Since in practice the numerical factor $M_R$ is likely to be of the order of $10^2 - 10^4$, and the other dimensionless parameters in eq.(24) are likely to be of the order of unity, the biomedical imaging quality of CT is likely to be much smaller than unity. Equation (24) can be used for design and optimization of PB-CT imaging setups, and for comparison with experiments, as demonstrated in the next section.

### 6. Experimental results

We carried out PB-CT imaging experiments at the Imaging and Medical Beamline (IMBL) of the Australian Synchrotron. We used a quasi-parallel X-ray beam with an energy of 32 keV ($\lambda$ = 0.3875 Å) and monochromaticity $\Delta\lambda / \lambda \approx 10^{-3}$. The source-to-sample and the sample-to-detector distances were respectively $R_1$ = 138 m and $R_2$ = 5 m. Therefore, the geometric magnification factor was $M = (R_1 + R_2) / R_1 \cong 1.036$ and the "effective defocus" distance was $R' = R_2 / M \cong 4.83$ m. We will explicitly apply below the simple corrections required for this "quasi-spherical" wave geometry (see e.g. Gureyev et al., 2008) to the results obtained in the previous sections for the parallel-beam geometry. In particular, the expression for the Fresnel number becomes $N_F = (\tilde{\Delta} / M)^2 / (\lambda R')$, where $\tilde{\Delta} / M$ is the spatial resolution "projected" from the image plane back to the object plane.

In order to be consistent with the first part of this paper (Gureyev et al., 2024), we use here a practical measure of the spatial resolution, Res, which is equal to twice the standard deviation of the corresponding PSF. For PSFs with a Gaussian and a rectangular shape, we get respectively (Gureyev et al., 2016):

$$\text{Res}[P_{Gauss}] = \pi^{-1/2} \Delta[P_{Gauss}] = \pi^{-1/2} \tilde{\Delta}[P_{Gauss}], \tag{25}$$

$$\text{Res}[P_{rect}] = (\pi / 3)^{-1/2} \Delta[P_{rect}] = \tilde{\Delta}[P_{rect}]. \tag{26}$$

A flat-panel (energy integrating) detector, Xineos 3030HR, with 99 μm × 99 μm pixels, and a photon-counting detector, Eiger2 3MW, with 75 μm × 75 μm pixels, were used in this experiment. The spatial resolutions, as measured in the collected projections, were $\text{Res[Xineos]} \cong 150.7\ \mu\text{m}$ and $\text{Res[Eiger]} \cong 86.2\ \mu\text{m}$ in the detector plane, which corresponded to $\tilde{\Delta}$ values in the object plane: $\tilde{\Delta}\text{[Xineos]} \cong \pi^{1/2} 150.7\ \mu\text{m} / 1.036 \cong 257.8\ \mu\text{m}$ and $\tilde{\Delta}[\text{Eiger}] \cong 86.2\,\mu\text{m}/1.036 \cong 83.2\,\mu\text{m}$. These calculations of $\tilde{\Delta}$ assumed a Gaussian-shape PSF for Xineos and a rectangular-shape PSF for Eiger. In the case of the Xineos detector, previous measurements produced similar values for Res values (Arhatari et al., 2021). In the case of the Eiger detector, the width of the rectangular PSF was supposed to be close to the pixel size of 75 μm (Dectris AG, 2025). However, that assumed that the lower energy threshold in the detector was set to a level ensuring complete rejection of split-pixel photon detection events. In the case of 32 keV X-rays, this threshold corresponds to 16 keV. However, in the present experiment we used the lower energy threshold setting of 4 keV in the Eiger detector, in order to increase the detective efficiency, even though it led to slightly worse spatial resolution. The resultant Fresnel numbers were respectively
$N_{F,M}[\text{Xineos}] = (257.8\,\mu\text{m})^2 / (4.83\,\text{m} \times 0.3875\,\text{Å}) \cong 355.1$ and
$N_{F,M}[\text{Eiger}] = (83.2\,\mu\text{m})^2 / (4.83\,\text{m} \times 0.3875\,\text{Å}) \cong 37.0$.

Two mastectomy samples from different donors were imaged in accordance with the Human Ethics Certificate of Approval and with written consent from each donor. The known average characteristics of adipose and glandular breast tissues for 32 keV X-rays implied that the ratio of the differences in real decrement to the differences in imaginary part of the refractive index of these two types of tissue was $\gamma \cong 869.4$ (TS-Imaging, 2025). Each sample was placed for imaging in a cylindrical thin-walled plastic container with a diameter of 11 cm and scanned in the PB-CT setup with the parameters listed above. Each scan contained 600 equispaced projections over 180 degrees rotation. By measuring the air kerma in the sample plane and using Monte Carlo simulations with breast-equivalent phantoms of the size corresponding to the imaged mastectomies, the incident X-ray flux and the rotation speed were adjusted to achieve the MGD of 4 mGy in each scan (Nesterets et al., 2015). Examples of projection images of the two samples with and without the TIE-Hom retrieval in accordance with eq.(4), $I_{retr}(\mathbf{r}_\perp, \theta) = (1 - a^2 \nabla_\perp^2)^{-1} I_R(\mathbf{r}_\perp, \theta)$, $a^2 = \gamma R' \lambda / (4\pi)$, are shown in Fig.1. Table 1 contains the results of measurements of the gain factor in the individual projections. According to the measurements, the average experimental gain factor in 2D projections was approximately 1.9 for Xineos and 4.2 for Eiger. The theoretical gain values, $G_2 = (\gamma / N_F)^{1/2}$, were respectively $G_2[\text{Xineos}] \cong (869.4/355.1)^{1/2} \cong 1.6$ and $G_2[\text{Eiger}] \cong (869.4/37.0)^{1/2} \cong 4.8$. The fact that these theoretical values were not exactly equal to the measured ones was due primarily to the uncertainty about the shape of PSFs of the detectors, especially in the case of the Xineos detector. Using the NRU, eq.(8), with the measured value $\tilde{\Delta}[\text{Eiger}] \cong 83.2\ \mu\text{m}$ and the ratio of average measured SNRs for Xineos and Eiger from Table 1, the value of $\tilde{\Delta}$ in the object plane for the Xineos detector can be estimated as $\tilde{\Delta}[\text{Xineos}] \cong 83.2\ \mu\text{m} \times (85.6/33.6) \cong 212.0\ \mu\text{m}$. The latter value is in between the values of $\tilde{\Delta}[\text{Xineos}]$ that are obtained for a rectangular and a Gaussian shape PSFs, given the measured value of $\text{Res}[\text{Xineos}] \cong 150.7\ \mu\text{m}/1.036 \cong 145.5\ \mu\text{m}$ in the object plane (see eqs.(25) and (26)). We will use the value of $\tilde{\Delta}[\text{Xineos}] \cong 212.0\ \mu\text{m}$ in the calculations below, which corresponds to $N_{F,M}[\text{Xineos}] = (212\,\mu\text{m})^2/(4.83\ \text{m} \times 0.3875\ \text{Å}) \cong 240.1$ and $G_2[\text{Xineos}] \cong (869.4/240.1)^{1/2} \cong 1.9$.

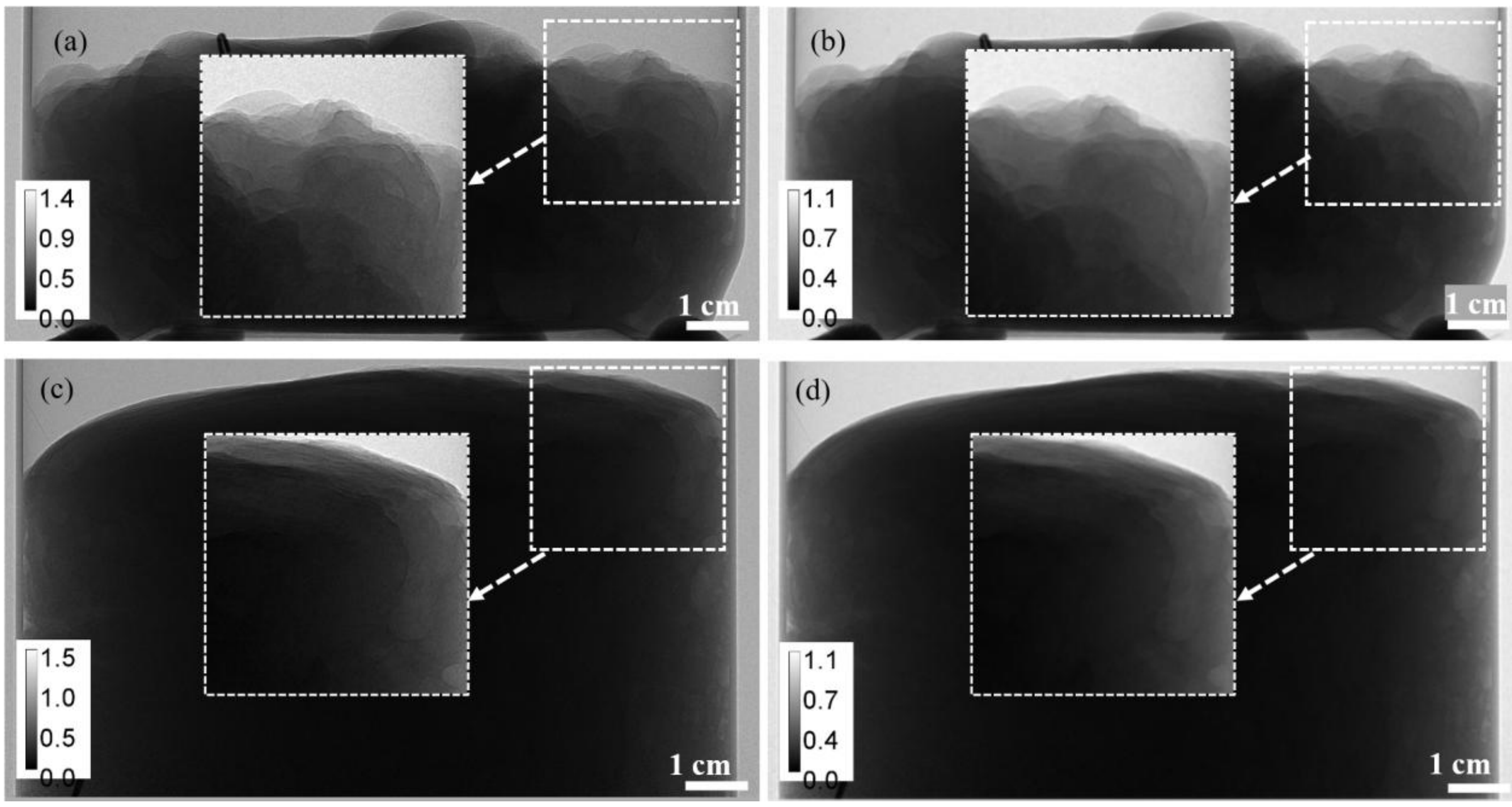

**Figure 1** PB-CT projections, normalized by flat-field images, of the two mastectomy samples collected at MGD ≅ 4 μGy, E = 32 keV and 5 m sample-to-detector distance, using the Eiger detector. (a) Sample 1, no TIE-Hom retrieval. (b) Sample 1, with TIE-Hom retrieval. (c) Sample 2, no TIE-Hom retrieval. (d) Sample 2, with TIE-Hom retrieval. Square inserts show magnified areas with fine details.

**Table 1** Measurements of SNR with and without the TIE-Hom retrieval in several PB-CT projections of two mastectomy samples.

| | SNR[$I_0$] | | SNR[$I_{retr}$] | | SNR[$I_{retr}$] / SNR[$I_0$] | |
|---|---|---|---|---|---|---|
| | Xineos | Eiger | Xineos | Eiger | Xineos | Eiger |
| Sample 1, proj.200 | 85.7 | 33.9 | 162.2 | 140.0 | 1.9 | 4.1 |
| Sample 1, proj.300 | 84.7 | 33.8 | 156.0 | 141.2 | 1.8 | 4.2 |
| Sample 1, proj.500 | 83.3 | 33.7 | 147.6 | 140.5 | 1.8 | 4.2 |
| | | | | | | |
| Sample 2, proj.200 | 86.8 | 33.5 | 168.3 | 140.0 | 1.9 | 4.2 |
| Sample 2, proj.300 | 85.6 | 33.3 | 163.9 | 139.0 | 1.9 | 4.2 |
| Sample 2, proj.500 | 87.6 | 33.5 | 171.8 | 143.0 | 2.0 | 4.3 |

The intrinsic imaging quality corresponding to the above imaging setup with the Xineos and Eiger detector can be evaluated according to eq.(9). Using the measured air kerma, we can calculate the incident fluence in the object plane: $I_{in,2D} = K_{3D,air}/(M_a R_{tr,air}) \cong 8\times10^{-3}\,\text{Gy}/(600\times7\times10^{-17}\,\text{Gy}\,\text{m}^2) \cong 0.19\ \mu\text{m}^{-2}$. Substituting the measured values of SNR and $\tilde{\Delta}$ for Xineos and Eiger, we obtain $\tilde{Q}_S[I_0, \text{Xineos}] \cong 85.6/(\sqrt{0.19\,\mu\text{m}^{-2}} \times 212.0\ \mu\text{m}) \cong 0.93$ and $\tilde{Q}_S[I_0, \text{Eiger}] \cong 33.6/(\sqrt{0.19\,\mu\text{m}^{-2}} \times 83.2\ \mu\text{m}) \cong 0.93$. With the TIE-Hom retrieval, the intrinsic imaging quality is increased to $\tilde{Q}_S[I_{retr}, \text{Xineos}] \cong 0.93\times1.9 \cong 1.77$ and $\tilde{Q}_S[I_{retr}, \text{Eiger}] \cong 0.93\times4.2 \cong 3.91$. We see, in particular, that the latter values are larger than unity, which would be impossible in conventional attenuation-based imaging.

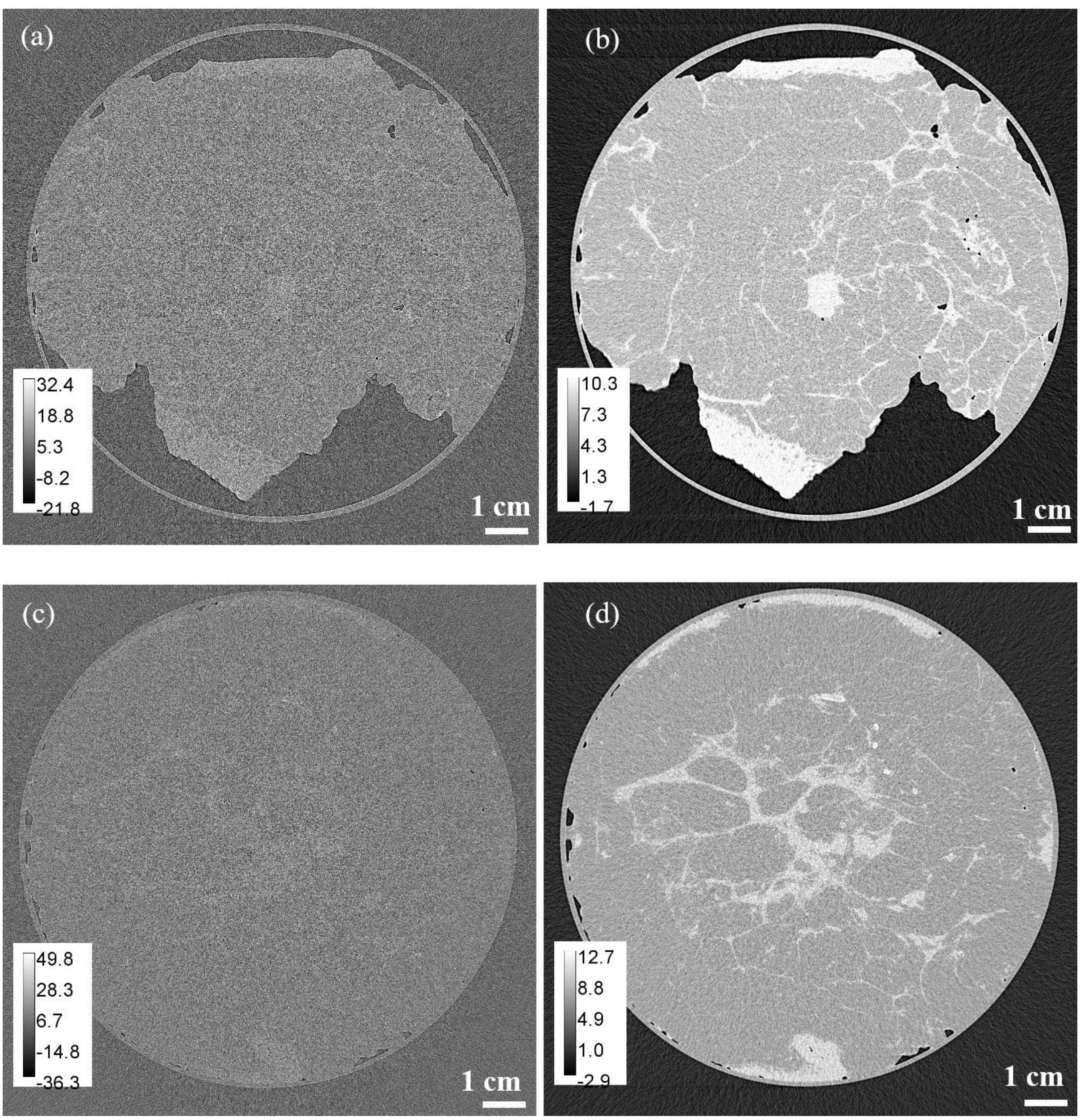

**Figure 2** PB-CT reconstructions (coronal slices) of the two mastectomy samples from a CT scan at MGD = 4 mGy, E = 32 keV and 5 m sample-to-detector distance, using the Eiger detector. (a) Sample 1, no TIE-Hom retrieval. (b) Sample 1, with TIE-Hom retrieval. (c) Sample 2, no TIE-Hom retrieval. (d) Sample 2, with TIE-Hom retrieval. These images contain the scaled distributions $\beta(\mathbf{r}) \times 10^{11}$.

**Table 2** Measurements of SNR with and without the TIE-Hom retrieval in several reconstructed coronal slices of two mastectomy samples.

| | SNR[$\beta_0$] | | SNR[$\beta_{retr}$] | | SNR[$\beta_{retr}$] / SNR[$\beta_0$] | |
|---|---|---|---|---|---|---|
| | Xineos | Eiger | Xineos | Eiger | Xineos | Eiger |
| Sample 1, slice 220 | 3.1 | 0.54 | 13.2 | 5.9 | 4.5 | 10.9 |
| Sample 1, slice 250 | 2.9 | 0.52 | 12.2 | 5.7 | 4.3 | 11.0 |
| Sample 1, slice 270 | 3.0 | 0.52 | 12.4 | 5.6 | 4.0 | 10.8 |
| | | | | | | |
| Sample 2, slice 220 | 2.8 | 0.51 | 11.8 | 5.5 | 4.2 | 10.7 |
| Sample 2, slice 250 | 2.7 | 0.51 | 11.7 | 5.6 | 4.3 | 10.9 |
| Sample 2, slice 350 | 2.7 | 0.53 | 11.9 | 5.6 | 4.3 | 10.7 |

We performed the reconstruction of "coronal" slices from CT scans collected using two different detectors, as described above. Examples of the reconstructed coronal slices of the two mastectomy samples are shown in Fig.2. The results of measurements of the SNR in the reconstructed slices and the 3D gain factors are presented in Table 2. The theoretical reasons for the measured values of SNR[$\beta$] in Table 2 being much smaller than the SNR[$I$] values in Table 1 are discussed in Appendix C. The results presented in the last two columns of Table 2 show that the SNR gain factor, compared to absorption-based imaging at the same dose and spatial resolution, was approximately 4.3 for the Xineos detector and 10.9 for the Eiger detector. The theoretical 3D gain factors in the present experiment, calculated in accordance with eq.(22) and eq.(23), were respectively 3.2 and 4.9 for the Xineos detector, and 9.5 and 11.6 for the Eiger detector. As can be seen here, the experimentally measured values of the 3D gain factor lied in between the theoretical estimates given by eqs.(22) and (23) for each of the two detectors.

While it was possible to perform the measurements of SNR[$I_0$] in the flat-field projection images, where the sample did not affect the measurements, a similar sample-independent measurement was impossible in 3D in the case of SNR[$\beta_0$]. Indeed, the mean reconstructed value (i.e. "the signal") in CT slices outside of the sample area is equal to zero in accordance with the formalism of CT reconstruction, since $\ln(I_0 / I_{in}) = 0$ in the absence of X-ray attenuation. Therefore, it was impossible to calculate sample-independent values of SNR and $\tilde{Q}_{S,3D}$ from the experimental data in this case. As measurements of SNR[$\beta_0$] presented in Table 2 already depend on the X-ray absorption in the sample, it makes more sense to instead calculate the 3D biomedical imaging quality characteristic $\tilde{Q}_{C,3D}$ in this case. The contrast between the mean values of $\beta$ in the glandular and adipose tissue in the reconstructed slices was $C_m[\beta] \cong (1.0 \times 10^{-10} - 7.3 \times 10^{-11}) / 1.0 \times 10^{-10} \cong 0.27$. The biomedical quality characteristic can be evaluated according to eq.(17a), $\tilde{Q}_{C,3D}[\beta_0] \cong CNR_{3D}\bar{R}_{ab,air}^{1/2}L^{1/2} / (\bar{D}_{g,3D}^{1/2}\tilde{\Delta}^{3/2})$, taking into account the MGD, $\bar{D}_{g,3D} = 4$ mGy, the radius of the cylinder containing the sample, $R = 5.5\text{ cm} = 2L/\pi$, and the value of $\bar{R}_{ab,air} \cong 7.003\times10^{-17}\text{ m}^4/\text{s}^2$ calculated above. Substituting all the relevant parameters into eq.(17a), we obtained $\tilde{Q}_{C,3D}[\beta_0,\text{Xineos}] \cong 9.7\times10^{-3}$ and $\tilde{Q}_{C,3D}[\beta_0,\text{Eiger}] \cong 7.2\times10^{-3}$, without the TIE-Hom retrieval, and $\tilde{Q}_{C,3D}[\beta_{retr},\text{Xineos}] \cong 4.1\times10^{-2}$ and $\tilde{Q}_{C,3D}[\beta_{retr},\text{Eiger}] \cong 7.8\times10^{-2}$, with the TIE-Hom retrieval. We see that the use of PB-CT substantially increased the biomedical imaging quality characteristic, in proportion to the 3D gain coefficients $G_3$ measured earlier (4.3 for the Xineos detector and 10.9 for the Eiger detector).

As explained above, the key reason for the values of $\tilde{Q}_{C,3D}[\beta_0]$ to be much smaller than unity is the mathematical ill-posedness of the CT reconstruction. If we substitute the relevant parameters of the present experiment, e.g. in the case of the Eiger detector, into eq.(24) with $G_3 = 1$, we obtain $\tilde{Q}_{C,3D} \cong 9.5\times10^{-3}$ (see Appendix C for details). This theoretical value is close to the value $\tilde{Q}_{C,3D}[\beta_0,\text{Eiger}] \cong 7.2\times10^{-3}$ measured in the experiment, confirming that in the present configuration $\tilde{Q}_{C,3D}[\beta_0]$ should indeed be expected to be much smaller than unity. As explained after eq.(24) and in Appendix C, the small value of $\tilde{Q}_{C,3D}[\beta_0]$ is determined by the degree of ill-posedness of the CT reconstruction in the present experimental configuration. The ill-posedness of the CT reconstruction is also behind the fact that while the $\tilde{Q}_S$ in 2D projections was found to be the same (equal to 0.93) for the Xineos and Eiger detectors, $\tilde{Q}_{C,3D}[\beta_0]$ was slightly higher for the Xineos detector ($9.7\times10^{-3}$) than for the Eiger detector ($7.2\times10^{-3}$). The reason for this "discrepancy" is that the squared ratio of the averaged SNR in the first and second columns of Table 2, $SNR^2[\beta_0,\text{Xineos}]/SNR^2[\beta_0,\text{Eiger}]$, was closer to the ratio of the fourth powers of the corresponding spatial resolutions, $\tilde{\Delta}^4[\text{Xineos}]/\tilde{\Delta}^4[\text{Eiger}]$, as expected in CT (Howells et al., 2009; Nesterets & Gureyev, 2014), rather than to the ratio of the third powers of the resolutions, $\tilde{\Delta}^3[\text{Xineos}]/\tilde{\Delta}^3[\text{Eiger}]$, as expected in the NRU and in the definition of the biomedical imaging quality.

**7. Conclusions**

We have analysed the performance of PB-CT in a general theoretical context using the NRU as the theoretical basis and the intrinsic imaging quality characteristic, $Q_S$, as a measure of system's performance. Recall that $Q_S^2$ is equal to the ratio of SNR$^2$ to the spatial resolution in the appropriate power $n$ that corresponds to the dimensionality of the images ($n = 2$ for projection imaging and $n = 3$ in the case of CT), normalized by the incident photon flux. According to NRU, $Q_S$ cannot be larger than unity or, more precisely, cannot be larger than the Epanechnikov constant, that is just slightly larger than one (de Hoog et al., 2014; Gureyev et al., 2016), for conventional absorption-based X-ray imaging systems. However, we have shown that in PBI setups the intrinsic imaging quality characteristic can be as high as $c_n(\gamma / N_F)^{n/4}$, where $\gamma$ is the ratio of the real decrement to the imaging part of the refractive index of the imaged object, $N_F$ is the minimal Fresnel number and $c_n$ is a dimensionless quantity of the order of one. This means that in PBI the maximum number of bits of information about the imaged sample delivered by each detected photon can be close to $(\gamma / N_F)^{n/4}$. In hard X-ray PBI of biological samples, the ratio $\gamma / N_F$ can be significantly larger than unity, which indicates potential for a beneficial violation of the NRU.

In order to objectively assess the improvement of biomedical image quality in PBI, one typically needs to evaluate the key image parameters, such as contrast, SNR and spatial resolution, as a function of not only the imaging setup, but also with respect to the radiation dose delivered to the sample in the process of imaging. The latter is particularly important in the case of medical X-ray imaging. For this purpose, in the present paper we introduced the biomedical imaging quality characteristic, $Q_C$, which was obtained by incorporating the contrast and radiation dose into the intrinsic imaging quality characteristic. Essentially, while $Q_S$ evaluates the objective performance of the imaging system alone, $Q_C$ allows one to assess the imaging quality of the system with respect to a particular sample or type of samples, by appropriately weighting the imaging quality metric in proportion to the contrast and in inverse proportion to the dose. We showed that the biomedical imaging quality in PB-CT is directly proportional to the image contrast and is maximized with respect to the X-ray absorption when the average transmission through the sample is equal to $1/e^2 \cong 0.135$. We also showed that the biomedical imaging quality in PB-CT is typically much smaller than unity because of the mathematical ill-posedness of the CT reconstruction (Natterer, 2001). We have previously compared objective assessments of imaging setups in terms of the measured $Q_S$ with the systematic subjective evaluation of the same images by groups of radiologists and medical imaging specialists (Baran et al., 2017; Donato et al, 2025; Pakzad et al., 2025). In future work, it will be important to also compare the (objective) biomedical imaging quality characteristic with the assessment of images by radiologists.

We have also analysed the results of an experiment conducted at IMBL (Australian Synchrotron) in which two full fresh mastectomy samples were imaged at 4 mGy MGD in PB-CT mode with 5 m sample-to-detector distance, using monochromatic parallel 32 keV X-rays and two different detectors. One of the detectors was a flat-panel detector with a pixel size of 99 μm and a spatial resolution of approximately 150 μm, while the other detector was a photon-counting one with 75 μm pixels and an approximately single-pixel PSF. We showed that the difference in the quality of the CT-reconstructed images obtained from the scans under the same conditions using the two different detectors was consistent with the theoretical results obtained earlier in this and related papers. In particular, the gain in SNR in PB-CT in comparison with conventional CT at the same dose and spatial resolution, was significantly larger in the 3D case, compared to the SNR gain in the individual 2D projections. In the case of the detector with 75 μm pixels, the gain in SNR in PB-CT was demonstrated to be larger than a factor of 10. Note that, since the radiation dose is proportional to $SNR^2$, a gain in SNR by a factor of 10, without loss of spatial resolution or contrast, corresponds to a potential reduction of the radiation dose by a factor of 100, without a loss of image quality. Even higher SNR gains (of the order of $10^2$) and dose reduction factors (of the order of $10^4$ - $10^5$), compared to equivalent conventional absorption-based imaging, were demonstrated previously (Kitchen et al., 2017) in a synchrotron-based small animal X-ray imaging experiment with a spatial resolution of approximately 40 μm. The present results, obtained for full intact human mastectomy samples under clinically acceptable radiation dose and spatial resolution, is an important step in our ongoing work towards breast cancer imaging of human patients at the Australian Synchrotron in the near future.

**Acknowledgements** We gratefully acknowledge the two donors of mastectomy samples, without whom this work would not have been possible. The experimental part of this research was undertaken on the Imaging and Medical beamline at the Australian Synchrotron, part of ANSTO.

## References

Arhatari, B.D., Stevenson, A.W., Abbey, B., Nesterets, Y.I., Maksimenko, A., Hall, C.J., Thompson, D., Mayo, S.C., Fiala, T., Quiney, H.M., Taba, S.T., Lewis, S.J., Brennan, P.C., Dimmock, M., Hausermann, D. & Gureyev, T.E. (2021). *Appl. Sci.* **11**, 4120.

Baran, P., Pacile, S., Nesterets, Y.I., Mayo, S.C., Dullin, C., Dreossi, D., Arfelli, F., Thompson, D., Lockie, D., McCormack, M., Taba, S.T., Brun, F., Pinamonti, M., Nickson, C., Hall, C., Dimmock, M., Zanconati, F., Cholewa, M., Quiney, H., Brennan, P.C., Tromba, G. & Gureyev, T.E. (2017). *Phys. Med. Biol.* **62**, 2315-2332.

Beltran, M.A., Paganin, D.M. & Pelliccia. D. (2018). *J. Opt.* **20,** 055605.

Bezak, E., Beddoe, A.H., Marcu, L.G., Ebert, M. & Price, R., Johns and Cunningham's The Physics of Radiology, 5th ed. (2021). (Charles C Thomas, Springfield).

Brombal L. (2020). *J. Instrum.* **15**, C01005.

Brombal, L., Donato, S., Dreossi, D., Arfelli, F., Bonazza, D., Oliva, P., Delogu, P., Di Trapani, V., Golosio, B., Mettivier, G., Rigon, L., Taibi, A. & Longo, R. (2018). *Phys. Med. Biol.* **63**, 24NT03.

Bronnikov, A. V. (1999). *Opt. Commun.* **171**, 239-244.

Dectris AG. (2025). Eiger2 detectors. https://www.dectris.com/en/detectors/x-ray-detectors/eiger2/eiger2-for-synchrotrons/eiger2-x-cdte/#Specifications (accessed 12 May 2025).

Donato, S., Caputo, S., Brombal, L., Golosio, B., Longo, R., Tromba, G., Agostino, R.G., Greco, G., Arhatari, B., Hall, C., Maksimenko, A., Hausermann, D., Lockie, D., Fox, J., Kumar, B., Lewis, S., Brennan, P.C., Quiney, H.M., Taba, S.T. & Gureyev, T.E. (2025). Accepted for publication in *Med. Phys.*

Goodman, P. (2000). Statistical Optics. (Wiley, New York).

Gureyev, T.E., Stevenson, A.W., Nesterets, Ya.I. & Wilkins, S.W. (2004). *Optics Commun.* **240**, 81-88.

Gureyev, T.E., Paganin, D.M., Myers, G.R., Nesterets, Ya.I. & Wilkins, S.W. (2006). *Appl. Phys. Lett.* **89**, 034102.

Gureyev, T.E., Nesterets, Y.I., Stevenson, A.W., Miller, P.R., Pogany, A. & Wilkins, S.W. (2008). *Opt. Expr.* **16**, 3223-3241.

Gureyev, T.E., Nesterets, Y.I., de Hoog, F., Schmalz, G., Mayo, S.C., Mohammadi, S. & Tromba, G. (2014). *Opt. Expr.* **22**, 9087-9094.

Gureyev, T. E., Nesterets, Y. I. & de Hoog, F. (2016). *Opt. Expr.* **24**, 17168–17182.

Gureyev, T. E., Nesterets, Y. I., Kozlov, A., Paganin, D. M. & Quiney, H. M. (2017). *J. Opt. Soc. Am. A* **34**, 2251–2260.

Gureyev, T. E., Kozlov, A., Nesterets, Y. I., Paganin, D. M., Martin, A. V. & Quiney, H. M. (2018). IUCrJ **5**, 716–726.

Gureyev, T.E., Kozlov, A., Nesterets, Y. I., Paganin, D. M. & Quiney, H. M. (2020). *Sci. Rep.* **10**, 7890.

Gureyev, T.E., Brown, H. G., Quiney, H.M. & Allen, L.J. (2022). *J. Opt. Soc. Am. A* **39**, C143-C155.

Gureyev, T.E., Paganin, D.M. & Quiney, H.M. (2024). *J. Synchrotron Rad.,* **31**, 896-909.

Higgins, P.D., Attix, F.H., Hubbell, J.H and Seltzer, S.W., Berger, M.J. & Sibata, C.H. (1992). Mass Energy-Transfer and Mass Energy-Absorption Coefficients, including In-Flight Positron Annihilation for Photon Energies 1 keV to 100 MeV, NISTIR 4812, NIST, Gaithesburg (https://nvlpubs.nist.gov/nistpubs/Legacy/IR/nistir4812.pdf)

Howells, M. R., Beetz, T., Chapman, H. N., Cui, C., Holton, J. M., Jacobsen, C. J., Kirz, J., Lima, E., Marchesini, S., Miao, H., Sayre, D., Shapiro, D. A., Spence, J. C. H. & Starodub, D. (2009). *J. Electron Spectros. Relat. Phenomena*, **170**, 4-12.

Hubbell, J.H. & Seltzer, S.W. (2004). X-Ray Mass Attenuation Coefficients, NIST Standard Reference Database 126. DOI: https://dx.doi.org/10.18434/T4D01F;

Johns, P.C. & Yaffe, M.J. (1985). *Med. Phys.* **12**, 289-296.

Kato, H. (2014). Japan. J. Radiolog. Technol. **70,** 684-691.

Kitchen, M. J., Buckley, G. A., Gureyev, T. E., Wallace, M. J., Andres-Thio, N., Uesugi, K., Yagi, N. & Hooper, S. B. (2017). *Sci. Rep.* **7**, 15953.

Liu, Q., Suleiman, M. E., McEntee, M. F. & Soh, B. P. *J. Radiol. Prot.* **42**, 011503 (2022).

Mandel, L. & Wolf, E. (1995). Optical Coherence and Quantum Optics (Cambridge University Press, Cambridge).

Marks, L.B., Yorke, E.D., Jackson, A., Ten Haken, R.K., Constine, L.S., Eisbruch, A., Bentzen, S.M., Nam, J. & Deasy, J.O. (2010). *Int. J. Radiat. Oncol. Biol. Phys.* **76**, S10–S19.

McEwen, B.F., Downing, K.H. & Glaeser, RM. (1995). *Ultramicroscopy* **60**, 357–373.

Momose, A. (2020). *Phys. Medica* **79**, 93-102.

Natterer, F. (2001). The Mathematics of Computerized Tomography. (SIAM, Philadelphia).

Nesterets, Y. I. & Gureyev, T. E. (2014). *J. Phys. D: Appl. Phys.* **47**, 105402.

Nesterets, Ya., Gureyev, T., Mayo, S., Stevenson, A., Thompson, D., Brown, J., Kitchen, M., Pavlov, K., Lockie, D., Brun, F. & Tromba, G. (2015). *J. Synchrotron Rad.* **22**, 1509-1523.

Nielsen, M. A., & Chuang, I. L. (2010). Quantum Computation and Quantum Information, 10th Ann. Ed. (Cambridge University Press, Cambridge).

Nugent, K.A., Paganin, D. & Gureyev, T.E. (2001). *Phys. Today* **54**, 27-32.

Paganin, D., Mayo, S. C., Gureyev, T. E., Miller, P. R. & Wilkins, S. W. (2002). *J. Microsc.* **206**, 33–40.

Paganin D., Gureyev T.E., Mayo S.C., Stevenson A.W., Nesterets Ya.I. & Wilkins, S.W. (2004). *J. Micros.* **214**, 315-327.

Pakzad, A., Turnbull, R., Mutch, S.J., Leatham, T.A., Lockie, D., Fox, J., Kumar, B., Häusermann, D., Hall, C.J., Maksimenko, A., Arhatari, B.D., Nesterets, Y.I., Entezam, A., Taba, S.T., Brennan, P.C., Gureyev, T.E. & Quiney, H.M. (2025). Eprint arXiv:2505.05812.

Ronneberger, O., Fischer, P. & Brox, T. (2015). U-Net: Convolutional Networks for Biomedical Image Segmentation in Medical Image Computing and Computer-Assisted Intervention -- MICCAI 2015, N. Navab, J. Hornegger, W. M. Wells, A. F. Frangi, Eds. (Springer International Publishing, 2015), pp. 234–241.

Sakurai, J.J. (1967). Advanced Quantum Mechanics (Addison-Wesley, Massachusetts), section 4.6.

Tavakoli Taba, S., Baran, P., Lewis, S., Heard, R., Pacile, S., Nesterets, Y. I., Mayo, S. C., Dullin, C., Dreossi, D., Arfelli, F., Thompson, D., McCormack, M., Alakhras, M., Brun, F., Pinamonti, M., Nickson, C., Hall, C., Zanconati, F., Lockie, D., Quiney, H. M., Tromba, G., Gureyev, T. E. & Brennan, P. C. (2019). *Acad. Radiol.* **26**, e79–e89.

Thompson, D.A., Nesterets, Y.I., Pavlov, K.M. & Gureyev, T.E. (2019). *J. Synchrotron. Rad.* **26**, 825-838.

TS-Imaging (2025). X-ray complex refraction coefficient calculator, http://ts-imaging.science.unimelb.edu.au/Services/Simple/ICUtilXdata.aspx (accessed 12 May 2025).

#### Appendix A. Calculation of the gain factor in PB-CT using NRU

Applying eq.(6) and then eq.(8) to the case of the filter function $O(\mathbf{r}) = T_{3,inv}(\mathbf{r})$, we can obtain:

$$G_3^2 = \frac{SNR^2[\beta_{retr}]}{SNR^2[\beta_0]} = \frac{SNR^2[T_{3,inv} * \mathbf{P}^{-1}C_R]}{SNR^2[\mathbf{P}^{-1}C_0]} = \frac{SNR^2[T_{3,inv} * \mathbf{P}^{-1}C_R]}{SNR^2[\mathbf{P}^{-1}C_R]} = \frac{\tilde{\Delta}^3[T_{3,inv} * \mathbf{P}^{-1}C_R]}{\tilde{\Delta}^3[\mathbf{P}^{-1}C_R]} . \quad \text{(A1)}$$

The first equality in eq.(A1) is the result of the assumption about having a fixed incident fluence and using “full” phase retrieval with $a = \sqrt{\delta R \lambda / (4\pi\beta)}$, as discussed in Section 5 in the paragraph following the one containing eq.(20). The second equality in eq.(A1) follows from eq.(6). The third equality in eq.(22) is based on the fact that, due to the photon flux conservation in free-space propagation, SNR$^2$ is the same in (flat regions of) CT-reconstructed volumes in the object and image planes: $SNR^2[\mathbf{P}^{-1}C_R] = SNR^2[\mathbf{P}^{-1}C_0]$ . Finally, the last equality in eq.(A1) is the consequence of eq.(8), with $O = T_{3,inv}$ and $I = \mathbf{P}^{-1}C_R$.

Using the definition of $\tilde{\Delta}$ and the fact that the convolution with $O = T_{3,inv}$ does not change the first integral norm, we can express the right-hand side of eq.(A1) as $\| \mathbf{P}^{-1}C_R \|_2^2 / \| T_{3,inv} * \mathbf{P}^{-1}C_R \|_2^2$. We assume that the width of the PSF in the image plane after the CT reconstruction is much narrower than $T_{3,inv}$, or equivalently, that the Fourier transform of $\mathbf{P}^{-1}C_R$ is much broader and is almost constant on the support of the Fourier transform of $T_{3,inv}$. Then using the Parseval theorem, we obtain:

$$\frac{\| T_{3,inv} * \mathbf{P}^{-1}C_R \|_2^2}{\| \mathbf{P}^{-1}C_R \|_2^2} \cong \frac{| \mathbf{P}^{-1}C_R |^2 (0)}{\| \mathbf{P}^{-1}C_R \|_2^2} 4\pi \int_0^{1/(2\tilde{\Delta}_{CT})} \frac{\rho^2 d\rho}{(1+4\pi^2 a^2 \rho^2)^2} = \frac{\tilde{\Delta}_{CT}^3}{8\pi a^3 f(A)} , \quad \text{(A2)}$$

where $\tilde{\Delta}_{CT}^3 \equiv \tilde{\Delta}^3[\mathbf{P}^{-1}C_R] = \| \mathbf{P}^{-1}C_R \|_1^2 / \| \mathbf{P}^{-1}C_R \|_2^2$, $f(A) = (\pi/2)[\arctan(A) - A/(1+A^2)]^{-1}$ and $A \equiv \pi a / \tilde{\Delta}_{CT}$. Note that $f(\infty) = 1$. From eq.(A1) and eq.(A2), we now obtain eq.(22).

#### Appendix B. Calculation of the gain factor in PB-CT using noise power spectrum

It is known (Goodman, 2000) that the noise variance is equal to the integral of noise power spectrum (NPS). In particular, $\sigma_{R,\beta}^2 = \int W_{R,\beta}(\mathbf{q}) d\mathbf{q}$, where $W_{R,\beta}(\mathbf{q})$ is the NPS of $\beta(\mathbf{r})$ reconstructed from a PB-CT scan collected at $z = R$. An explicit expression for $W_{R,\beta}(\mathbf{q})$ can be obtained in conjunction with eq.(6), in the case of a PB-CT scan with $M_a$ projections uniformly distributed over the 180 degrees of sample rotation and uncorrelated photon fluence with $\overline{n}_{det}$ photons on average per detector pixel (Nesterets & Gureyev, 2014):

$$W_{R,\beta}(\rho,\eta,\theta) = \frac{\pi\,\rho\,\tilde{\Delta}_{pix}^2 \chi(\rho,\eta)}{(2k)^2 M_a \overline{n}_{det}[1+4\pi^2 a^2(\rho^2+\eta^2)]^2}, \tag{B1}$$

where $\tilde{\Delta}_{pix}$ is the linear size of the (square) detector pixels and we assumed for simplicity that the detector has a rectangular-shape modulation transfer function $\chi$ with the width equal to $1/\tilde{\Delta}$, where $\tilde{\Delta}$ is the spatial resolution. Integrating eq.(B1) with respect to $\rho$ and $\eta$ over the range of all detectable frequencies, we obtain:

$$\sigma_{R,\beta}^2 = \frac{\tilde{\Delta}_{pix}^2}{16\pi^2(2k)^2 a^4 \overline{n}_a} \int_{-\pi a/\tilde{\Delta}}^{\pi a/\tilde{\Delta}} \int_{-\pi a/\tilde{\Delta}}^{\pi a/\tilde{\Delta}} \frac{u^2 du dv}{(1+u^2+v^2)^2}, \tag{B2}$$

where $\overline{n}_a \equiv M_a \overline{n}_{det}$ is the mean number of photons registered in one detector pixel during the CT scan. Unfortunately, we are not aware of a method that would allow one to calculate the double-integral in eq.(B2) exactly analytically. Therefore, the following results rely on approximations, similar to the results in (Nesterets & Gureyev, 2014). The behaviour of eq.(B2) is determined primarily by the ratio $a/\tilde{\Delta} = [\gamma/(4\pi N_F)]^{1/2}$. When $a/\tilde{\Delta} << 1$, the denominator in eq.(B2) can be approximated by 1, and the simple integration results in

$$\sigma_{0,\beta}^2 \cong \frac{\pi^2 \tilde{\Delta}_{pix}^2}{12(2k)^2 \overline{n}_a \tilde{\Delta}^4}. \tag{B3}$$

The latter result becomes exact in the case of $a = 0$, i.e. in the case of conventional CT with $R = 0$. On the other hand, when $a/\tilde{\Delta} >> 1$, eq.(B2) can be approximated as

$$\sigma_{R,\beta}^2 \cong \frac{[\ln(\gamma/N_F)-1]\tilde{\Delta}_{pix}^2}{32\pi(2k)^2 \overline{n}_a a^4}. \tag{B4}$$

Taking the square root of the ratio of eq.(B4) and eq.(B3), we obtain eq.(23).

**Appendix C. Biomedical imaging quality in CT**

Here we investigate why the measured values of $SNR[\beta_0]$ in Table 2 were significantly lower that the measured values of $SNR[I]$ in Table 1. A related question is about the measured values of $\tilde{Q}_{C,3D}[\beta_0]$ being significantly smaller than unity. It can be shown from eq.(B3), or from the results in (Gureyev et al., 2016), that

$$SNR^2[\beta_0] \cong (12/\pi^2)\bar{n}_a \bar{\mu}^2 \tilde{\Delta}^4 / \tilde{\Delta}^2_{pix}, \qquad \text{(C1)}$$

where $\bar{\mu} = 2k\bar{\beta}_0$. Note that the mean number of photons collected in a single detector pixel with the size $\tilde{\Delta}_{pix}$ during the CT scan is $\bar{n}_a = M_a I_{in,2D} \exp(-\bar{\mu}L)\tilde{\Delta}^2_{pix}$. Equation (C1) implies, in particular, that $SNR[\beta_0]$ is proportional to the fourth power of the spatial resolution, as mentioned earlier in the paper:

$$SNR^2[\beta_0] \cong (12/\pi^2)\bar{\mu}^2 \exp(-\bar{\mu}L) M_a I_{in,2D} \tilde{\Delta}^4 . \qquad \text{(C2)}$$

In the experiment considered in Section 6, the effective spatial resolution of the Eiger detector was $\tilde{\Delta} = 83.2\ \mu\text{m}$, $M_a = 600$ and the 2D photon fluence was $I_{in,2D} \cong 0.19\ \mu\text{m}^{-2}$. Using the known values for the linear absorption coefficient of adipose tissue at 32 keV, $\mu \cong 2.62\times10^{-5}\ \mu\text{m}^{-1}$ (TS-Imaging, 2025), we get $\bar{\mu}L \cong 2.62\times10^{-5}\ \mu\text{m}^{-1}\times 8.64\ \text{cm} \cong 2.26$ .It then follows from eq.(C2) that $SNR[\beta_0] \cong 0.69$. This theoretical value is close to the average measured value $SNR[\beta_0] \cong 0.52$ for the Eiger detector in Table 2 and is also substantially lower than the values of $SNR[I]$ in Table 1. The last fact does not have a fundamental nature, because, at a given radiation dose, the ratio of SNRs in the CT projections and in the reconstructed slices of $\beta$ depends on the number of projections in the CT scan.

Substituting eq.(C2) and the expression for the MGD $\bar{D}_g \cong D^N_g M_a \bar{I}_{in,2D} \bar{R}_{ab,air}$ into eq.(17a), we obtain

$$\tilde{Q}^2_{C,3D}[\beta_0] = \frac{12C^2_m (\bar{\mu}L)^2 \exp(-\bar{\mu}L)\tilde{\Delta}}{\pi^2 D^N_g L} = \frac{12C^2_m (\bar{\mu}L)^2 \exp(-\bar{\mu}L)}{\pi^2 D^N_g M_R}, \qquad \text{(C3)}$$

where $M_R = \pi R_C / (2\tilde{\Delta})$ is equal to π / 2 times the number of spatial resolution units in the radius of the CT reconstruction cylinder. In the experiment described in Section 6, we had $C_m \cong 0.27$, $\bar{\mu}L \cong 2.26$, $D_g^N \cong 0.5$ and $M_R = \pi \times 5.5\,\text{cm}/(2\times 83.2\,\mu\text{m}) \cong 1{,}038$. Substituting these numbers into eq.(C3), we obtain $\tilde{Q}_{C,3D} \cong 9.5\times 10^{-3}$. The latter theoretical value is close to the value $\tilde{Q}_{C,3D}[\beta_0,\text{Eiger}] \cong 7.2\times 10^{-3}$ measured in the experiment in Section 6. This theoretical estimation confirms that $\tilde{Q}_{C,3D}[\beta_0]$ in the present configuration should indeed be expected to be much smaller than unity. It was already noted in Section 4 that $\tilde{Q}_{C,3D} \sim 1$ requires $SNR \sim 10^3$, while in the present configuration we had $SNR[\beta_0] \cong 0.52$. The key reason for the relatively small value of $\tilde{Q}_{C,3D}$ in CT reconstruction is the ill-posedness of the CT reconstruction, which leads to the appearance of the (typically, large) numerical factor $M_R$ in the denominator of eq.(C3). At the same time, the term $(\bar{\mu}L)^2 \exp(-\bar{\mu}L)$ in the numerator of eq.(C3) has the theoretical maximum equal to approximately 0.54, which is achieved at $\bar{\mu}L = 2$ (the latter value is quite close to the value $\bar{\mu}L \cong 2.26$ in our experiment described in Section 6). Note that the SNR for "projected" values of $\beta$, e.g. for $\beta$ values integrated or averaged along one of the spatial coordinates, can be several orders of magnitude higher than for the 3D distribution of $\beta$. In other words, the biomedical imaging quality of the synthetic 2D mammographic images, obtained from CT-reconstructed slices, can be close to unity. This shows that the "dose fractionation theorem" (Howells et al., 2009; McEwen et al., 1995) does not hold for the types of CT imaging considered in the present paper, which is again explained by the mathematical ill-posedness of CT reconstruction; see also a related discussion in (Gureyev et al., 2018).